\documentclass[superscriptaddress,pra,aps,twocolumn,10pt,longbibliography]{revtex4-1}
\usepackage{bm}
\usepackage{amsmath}
\usepackage{amsfonts}
\usepackage{graphicx}
\usepackage{grffile}
\usepackage{natbib}
\usepackage{epstopdf}
\usepackage[pdftex]{color}
\usepackage[dvipsnames]{xcolor}
\usepackage{hyperref}
\usepackage{verbatim}
\usepackage{float}
\usepackage[english]{babel}
\usepackage[boxed]{algorithm}
\usepackage{algpseudocode}
\hypersetup{
    colorlinks=true,       % false: boxed links; true: colored links
    linkcolor=cyan,          % color of internal links
    citecolor=magenta,        % color of links to bibliography
    filecolor=magenta,      % color of file links
    urlcolor=cyan,           % color of external links
    runcolor=cyan
}

\newcommand{\beq}{\begin{eqnarray}}
\newcommand{\eeq}{\end{eqnarray}}
\newcommand{\bes} {\begin{subequations}}
\newcommand{\ees} {\end{subequations}}

\newcommand{\ignore}[1]{}

\graphicspath{{figs/}}

\begin{document}

\title{Perils of Embedding for Sampling Problems}

\author{Jeffrey Marshall}
\affiliation{QuAIL, NASA Ames Research Center, Moffett Field, California 94035, USA}
\affiliation{USRA Research Institute for Advanced Computer Science, Mountain View, California 94043, USA}

\author{Andrea Di Gioacchino}
\affiliation{Dipartimento di Fisica, University of Milan and INFN, via Celoria 16, 20133 Milan, Italy}

\author{Eleanor G. Rieffel}
\affiliation{QuAIL, NASA Ames Research Center, Moffett Field, California 94035, USA}

\begin{abstract}
Advances in techniques for thermal sampling in classical and
quantum  systems  would  deepen  understanding  of  the  underlying  physics. Unfortunately, one 
often has to rely solely on inexact numerical simulation, due to the intractability of computing the
partition  function  in  many  systems  of  interest. Emerging  hardware,  such  as  quantum  annealers,
provide novel tools for such investigations, but it is well known that studying general, non-native systems
on such devices requires graph minor embedding, at the expense of introducing additional variables. 
The effect of embedding for sampling is more pronounced than for optimization; for optimization one is just concerned with the ground state physics, whereas for sampling one needs to consider states at all energies. 
We argue that as the system size or the embedding size grows, the chance of a sample being in the
subspace of interest –- the logical subspace -– can be exponentially suppressed.  Though the severity of this
scaling  can  be lessened  through  favorable  parameter  choices, certain physical constraints (such as a fixed temperature and range of couplings) provide hard limits on what is currently feasible.
Furthermore, we show that up to some practical and reasonable assumptions, any type of post-processing to project samples back into the logical subspace will bias the resulting statistics. We introduce a new such technique, based on resampling, that substantially outperforms majority vote, which is shown to fail quite dramatically at preserving distribution properties.
\end{abstract}
\maketitle

\section{Introduction}
Improving the efficiency of sampling from certain distributions, such as Boltzmann distributions, could provide significant benefits for understanding equilibrium physics of many body systems, phase transitions in spin glasses, and for certain practical applications in the fields of
machine learning and optimization. Sampling is a challenging task; for example, sampling from a Boltzmann distribution at sufficiently cold  temperature is NP-hard. Special-purpose hardware, such as quantum annealers, have been proposed as potentially providing improved sampling capabilities, for tasks including machine learning, and physics simulation \cite{adachi,perdomo,DW2x-gibbs-ML,Amin:boltzmann,qvae,QAAAN,boltzmann-li-qac,boltzmann-galaxy,harris-tfim}. 

Many interesting cases can be reduced to sampling from a Boltzmann distribution $e^{-\beta H(\boldsymbol{s})}/Z$ with $H(\boldsymbol{s})$ a classical Ising model Hamiltonian of the form
\begin{equation}
    \label{eq:ising}
    H(  \boldsymbol{s} ) = \sum_{i,j}J_{ij}s_i s_j + \sum_i h_i s_i,
\end{equation}
where the real-valued \textit{couplings} $J_{ij}$ and \textit{local fields} $h_i$ fully specify the problem, with the partition function $Z=\sum_{\boldsymbol{s}} \exp(-\beta H(\boldsymbol{s}))$ for normalization of the probability distribution.
The energy (cost) associated with state 
$\boldsymbol{s} = (s_1,\dots , s_N )$ is given by $ H(  \boldsymbol{s} )$, where the spin variables $s_i$ take values in $\{-1,1\}$.
For optimization purposes, one is interested in the low cost configurations, or ideally the global minimum.

\begin{figure}
    \centering
    \includegraphics[width=0.95\columnwidth]{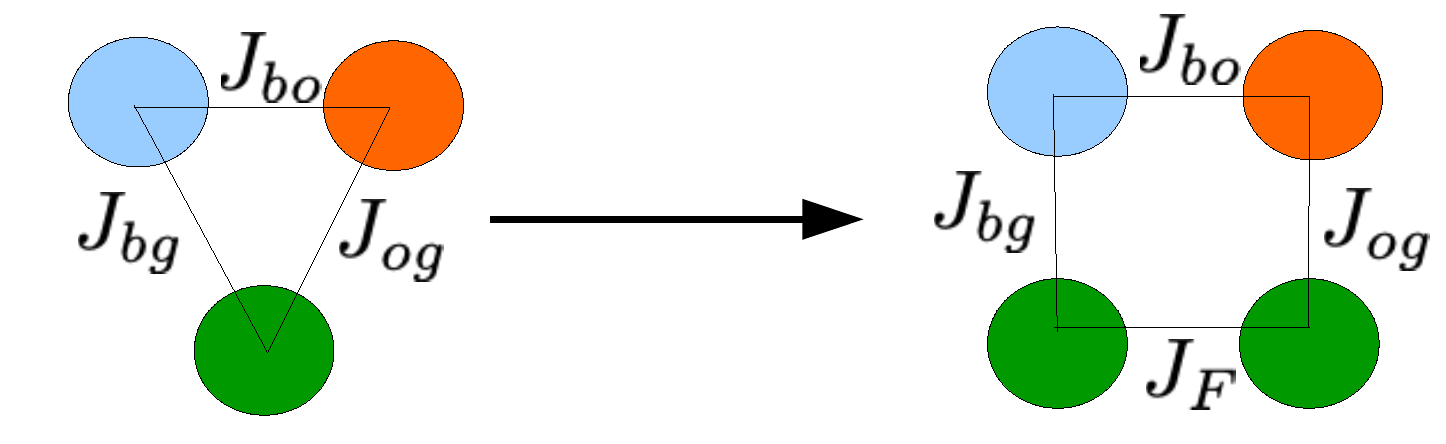}
    \caption{Example of embedding a fully connected graph of three nodes (triangle) into a square graph. The edges have weights given by  $J_{ij}$ defined by the Hamiltonian Eq.~(\ref{eq:ising}).
    In the embedded graph (right) an additional variable is used, with the green vertex being split into two, coupled with strength $J_F$. This combined variable is often referred to as a logical vertex, or logical spin. Note, the embedding process is in general not unique.}
    \label{fig:embedding_example}
\end{figure}

Depending on the problem one is considering, the couplings can define  a complicated graph, such as a 3-dimensional graph, or even a fully connected graph. 
Hardware constraints restrict the class of Hamiltonians that can be natively implemented on certain emerging hardware, including the D-Wave quantum annealing devices, and other special purpose Ising machines \cite{johnson:11,ising-machine-QbM}. General Hamiltonians can be mapped to native Hamiltonians, but care must be taken to understand what properties carry over and which do not.
A common constraint is that in superconducting qubit processors, including both universal processors and quantum annealers, 
only select couplings are available, often just between nearest neighbor qubits on the chip. 
To overcome connectivity limitations, {\it minor embedding} is used, mapping Eq.~(\ref{eq:ising}) to a new Hamiltonian of a similar form, but with only couplings 
%are only specified over 
native to the hardware graph
\begin{equation}
\label{eq:ising-mapped}
    \tilde{H}(  \tilde{\boldsymbol{s}}) = \sum_{\langle i,j\rangle} \tilde{J}_{ij}\tilde{s}_i \tilde{s}_j + \sum_i \tilde{h}_i \tilde{s}_i,
\end{equation}
where angle brackets 
indicate the sum is over the restricted graph given by the hardware, and 
$\tilde{\boldsymbol{s}}$ necessarily contains more variables than $  \boldsymbol{s}$. See Fig.~\ref{fig:embedding_example} for a simple example.
For optimization, the requirement on embedding is that from the global minimum of Eq.~(\ref{eq:ising-mapped}) one can infer the global minimum of Eq.~(\ref{eq:ising}). We call this the \textit{global-to-global} property.
Embedding, and the related topic of \textit{parameter setting}, is a well studied concept, 
beginning with early work of Choi \cite{Choi1,Choi2}.

To isolate the issues introduced by minor embedding from other implementation issues that may bias the sampling, we consider the following abstract problem. Suppose that one is interested in sampling a thermal distribution for $H$, but can only receive samples from a thermal distribution for $\tilde H$, a minor embedding of $H$. To what extent can we sample from $H$'s thermal distribution using samples from $\tilde H$'s distribution? While this problem is motivated in part by quantum annealing where minor embedding is a standard tool, the problem is a purely classical physics problem. In particular, our analysis is agnostic as to whether the thermal samples for $\tilde H$ are obtained from classical or quantum hardware.
As an aside, we remark that there are generalizations of this problem to quantum Hamiltonians, but we consider only classical Hamiltonians here.

More specifically, imagine the goal is to sample from a (thermal) distribution $\mathcal{D}$ which depends on Hamiltonian $H$ Eq.~(\ref{eq:ising}), obtaining samples of the form $\mathcal{D}(H)$.
If one has a perfect sampler, but with a restricted topology, instead one will sample from a Hamiltonian $\tilde{H}$ of the form Eq.~(\ref{eq:ising-mapped}), thus obtaining samples $\mathcal{D} ( \tilde{H})$.
In order to sample from the target Hamiltonian Eq.~(\ref{eq:ising}), one therefore needs to perform a projection on the sampled distribution $\Pi : \mathcal{D}(\tilde{H}) \rightarrow \mathcal{D}(H) $.
The goal is to find a suitable choice of $\Pi$ so that the target distribution is faithfully represented.

We focus on the case where $\mathcal{D}$ corresponds to a Boltzmann distribution, i.e. $\mathcal{D}(H) = \exp(-\beta H)/Z$ where $Z$ is the partition function, and $\beta$ an inverse temperature.
These distributions are of particular relevance 
given recent work using quantum annealers with a restricted topology to generate thermal samples from (classical) Hamiltonians of the form Eq.~(\ref{eq:ising}), for use in machine learning \cite{adachi,perdomo,DW2x-gibbs-ML,Amin:boltzmann,qvae,QAAAN,boltzmann-li-qac,boltzmann-galaxy}.
The main goal of this work is to demonstrate that as system sizes increase, the greater the need to develop new techniques for mapping from the embedded distribution to the native, logical distribution.
Our results are three fold.
First we will outline in more detail the problem of sampling from an embedded problem. In particular, we argue, and demonstrate numerically for small scale systems, that the number of samples received from $\mathcal{D}(\tilde{H})$ requiring a non-trivial projection procedure can grow exponentially in system size $N$. That is, the probability of observing a sample from within the logical subspace can decrease exponentially, provided there are constraints such as fixed maximum coupling values and temperature.
We also show that, under a reasonable set of assumptions, it is not possible to find a  projection $\Pi$ that preserves Boltzmann distributions exactly.
To highlight this further, we study perhaps the simplest (and most common) type of projection technique, typically referred to as majority vote (MV), showing that it is a poor choice in general. 
Next, we introduce a resampling technique (that we call RRS), which empirically outperforms MV. We finish with a discussion and outline possible future research directions based on this work.

\section{Embedding: definitions and nomenclature \label{sect:embedding}}

A minor embedding (henceforth, just `embedding')  uses
multiple physical spins (vertices), and couplings between them, to represent single spins in the original problem on the connectivity-limited hardware.
If one performs an edge contraction over these
vertices in a specified manner, one will arrive at the 
graph for the original Hamiltonian.
This general idea is illustrated in Fig.~\ref{fig:embedding_example}, where a triangular graph is embedded into a square graph, resulting in one additional variable, and one additional coupling which we denote by $J_F$. The task of picking $J_F$ requires special attention; lower bounds on choices of the additional parameters to achieve the global-to-global property are given in Refs.~\cite{Choi1,Choi2}.

More formally, consider the graph $G_H$ associated with Hamiltonian $H$ of the form Eq.~(\ref{eq:ising}). Each spin $s_i$ in the model $H$ defines a vertex $i$ in $G_H$, and a coupling between spins $J_{ij}$ defines a weighted undirected edge between vertex $i$ and $j$. Each node also has associated with it the corresponding local field $h_i$.

The graph $G_H$ is embeddable in another graph $\tilde{G}$ if there exists a mapping $\phi : G_H \rightarrow \tilde{G}$ such that 1) each node $i$ of $G_H$ is mapped to a (connected) subtree $T_i$ of $\tilde{G}$, with $T_i\cap T_j = \emptyset$ for $i\ne j$, and 2) for each edge $(i,j)$ of $G_H$ of weight $J_{ij}$, there are edges from $T_i$ to $T_j$ in $\tilde{G}$ which cumulatively sum to $J_{ij}$.
We also require that the local fields of each $T_i$ sum to $h_i$.
In this way, $G_H$ can be constructed from $\tilde{G}$ by contracting the edges of each $T_i$, i.e. $G_H$ is a graph minor of $\tilde{G}$.
Since the subtrees $T_i$ necessarily introduce additional variables, the dimensionality of the configuration space $\mathrm{dim} \tilde{\mathcal{H}} = 2^{\tilde{N}} \ge \mathrm{dim} \mathcal{H} = 2^N$ where $\tilde{\mathcal{H}}$ and $\mathcal{H}$ are the configuration spaces for the models $\tilde{H}$ and $H$ respectively, with $\tilde{N}$ and $N$ variables.

A configuration $  \tilde{\boldsymbol{s}}_L \in \tilde{\mathcal{H}}$, for which in each subtree $T_i$ the spins are all aligned identically, is known as a logical configuration, and belongs to the logical subspace $\tilde{\mathcal{H}}_L \subset \tilde{\mathcal{H}}$ of size $\mathrm{dim} \tilde{\mathcal{H}}_L = 2^N$. 
Any configuration in $\tilde{\mathcal{H}}_L$ has a corresponding and unique configuration in $\mathcal{H}$ which is found by simply replacing the identically pointing spins in each subtree by a single spin of same orientation.
We will therefore throughout refer to the subtrees $\{T_i\}_{i=1}^N$ as \textit{logical subtrees}, or as \textit{logical spins} when referring to the equivalent variables in model $\tilde{H}$.
If a logical spin contains spins of differing orientations, we will often refer to these as \textit{broken}.

In order to encourage the spins composing a logical spin to align under thermal sampling, strong ferromagnetic bonds $J_F < 0$ can be placed between the vertices in the logical subtree, so that there is a cost penalty related to $|J_F |$ whenever a spin is misaligned. If $J_F$ can be chosen to be infinitely large and negative, thermodynamic sampling at finite temperature guarantees one never observes a configuration outside of the logical space.
Practically, however, the size of $|J_F|$ is limited, both by the hardware, and since too large a $|J_F|$ can introduce large energy barriers and deep local minima in the landscape of the problem, making it prohibitive for thermal -- including non-zero temperature quantum annealing -- algorithms to traverse. We discuss these points in more detail in Sect.~III.

Embeddings of this type guarantee that for any configuration $  \boldsymbol{s} \in \mathcal{H}$ of cost $H(  \boldsymbol{s})$, there is an equivalent logical configuration $ \tilde{\boldsymbol{s}} \in \tilde{\mathcal{H}}_L$  with cost $\tilde{H}( \tilde{\boldsymbol{s}}) = H(  \boldsymbol{s}) + C$ where $C$ is a constant and global energy shift (i.e. independent of any particular $ \boldsymbol{s}$).
If subtree $T_i$ in $\tilde{G}$ contains $n_i$ vertices, with edge weights all $J_F$, the energy shift $C$ is simply given by 
\begin{equation}
    C = J_F \sum_{i=1}^N (n_i-1).
\end{equation}
This property is crucial for sampling purposes since it guarantees relative thermal sampling weights $w_{ij} := \exp (-\beta (H(\boldsymbol{s}_i) - H(\boldsymbol{s}_j)))$ are preserved by the embedding process, where $\boldsymbol{s}_{i,j}$ are spin configurations. 
In particular, if we denote the Boltzmann distribution for Hamiltonian $H$ at inverse temperature $\beta$ over  $\mathcal{H}$  by $\mathcal{D}_{\mathcal{H}}(H,\beta)$, then, restricting to the logical subspace of the embedded problem preserves the distribution: 
\begin{equation}
    \mathcal{D}_{\tilde{\mathcal{H}}_L} (\tilde{H},\beta) = \mathcal{D}_{\mathcal{H}}(H,\beta).
    \label{eq:dist}
\end{equation}

\subsection{Embedding Graph \label{sect:embedding_graph}}
Throughout this work, we use as our hardware restricted graph $\tilde{G} = \tilde{G}(K,J_F,N)$ one in which each subtree is a \textit{chain} (i.e.~a path) with the same number of vertices $K$, and internal logical spin couplings all of the same strength $J_F$. The total number of spins is $N\times K$. Each problem coupling $J_{ij}$ of $H$ is a single edge in $\tilde{G}$ also of weight $J_{ij}$, and local fields, $h_i$, are divided evenly between each spin in a logical spin (i.e.~with value $h_i/K$).

In the hardware graph, each spin has coordinate $(k,i)$ where $i$ is the logical spin index (equivalent to a vertex index in $G(H)$), and $0\le k<K$ denoting the spins position within the chain.
We have two ways to connect logical spins in the hardware graph. 
If there is an edge $J_{ij}\neq 0$ in $G(H)$, we can either i) pick random $0\le k_i,k_j <K$ such that there is an edge $((k_i,i),(k_j,j))$ with weight $J_{ij}$ in $\tilde{G}$, or ii) follow a deterministic embedding such that:
for $j>i$, vertex $(k,i)$ connects to $(k,j)$ if $j = i + (K-k) + nK$ for $n=0,1,\dots$, with weight $J_{ij}$.

This flexibility allows us to either i) simulate random embeddings in the hardware graph, or, ii) perform a direct comparison between different problems using a fixed embedding procedure.
The first point is intended to address the fact, as mentioned in Sect.~\ref{sect:embedding}, that there is typically not a unique choice of embedding, and the second point is so we can later compare between different projection techniques using the same embedding.

\begin{figure}
    \centering
    \includegraphics[width=0.7\columnwidth]{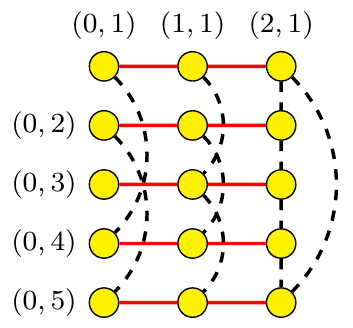}
    \caption{Example of our embedding graph $\tilde{G}(K=3,J_F,N=5)$ for a $5$ variable fully connected graph. Each horizontal row of three spins (yellow circles) is a logical spin (subtree); each variable is represented by  $K=3$ logical variables in this example. Some labels with spin coordinates are shown: each spin has coordinate $(k,i)$, where $i$ is the logical spin index and $k$ denotes the spin position within the chain. Red (solid) lines indicate ferromagnetic couplers of strength $J_F<0$ `gluing' the logical spins, and the black (dash) lines are the problem couplings $J_{ij}$ between variables. Local fields are also present (divided evenly across the spins in a subtree), but are not shown for simplicity. For larger problems, or different logical subtree sizes ($K$), this basic structure can be repeated indefinitely (see main text). Performing an edge contraction over the red (solid) ferromagnetic edges results in the native fully connected problem.}
    \label{fig:fc_cartoon}
\end{figure}

An example of our hardware graph is shown in Fig.~\ref{fig:fc_cartoon}, for $K=3$, for the deterministic embedding.

We pick this graph $\tilde{G}$ since each logical spin is treated equivalently, therefore allowing us to study directly the effect of changing $J_F$ and $K$ on sampling quality. Moreover, we can embed any type (i.e. fully connected) of graph of size $N$ into $\tilde{G}(K,J_F,N)$. Throughout, our units are defined relative to the native Hamiltonian, i.e. relative to $\max \{|J_{ij}|,|h_i|\}$ (which we pick here to be 1 for convenience).

\section{The problem of {sampling after embedding: analytical results}}\label{sect:problem_embedding}

Sect.~\ref{sect:embedding} introduced the key ideas behind embedding. We will now elaborate on this to highlight potential issues using embeddings in a sampling task. We focus on the task of Boltzmann sampling, however similar arguments can be applied to any form of sampling in which the statistics may be biased by embedding and projecting.

Our main result is an equation which shows that for a given embedding, and at fixed temperature, the probability of observing a configuration within the logical subspace $\tilde{\mathcal{H}}_L$ is exponentially small in problem size $N$, and also the subtree sizes $K$.  This means that it is not practical to simply restrict to this subspace and utilize Eq.~(\ref{eq:dist}).

\begin{figure}
    \centering
    \includegraphics[width=0.95\columnwidth]{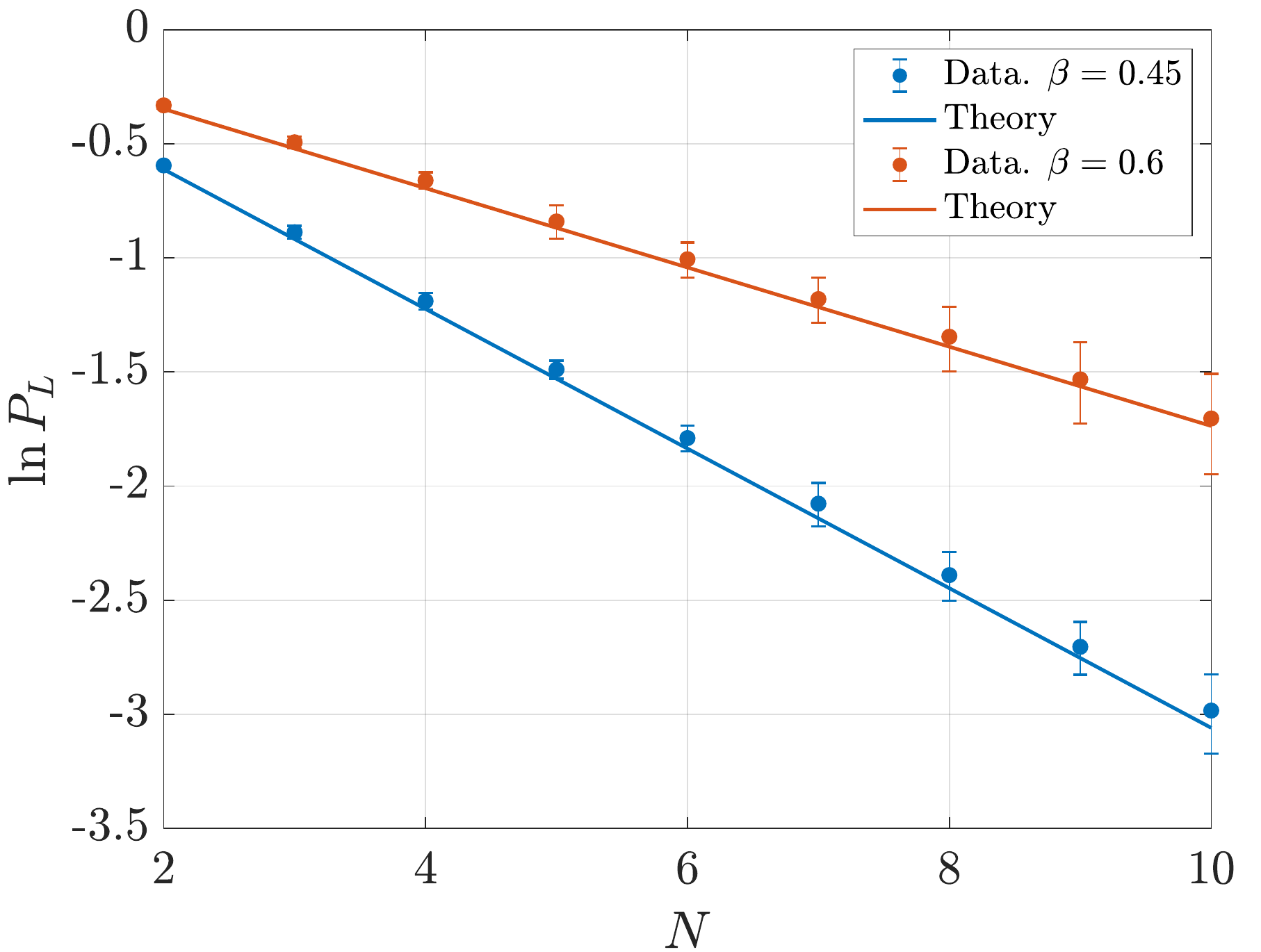}
    \caption{The probability $P_L\equiv P_0$ of observing a configuration from the logical subspace under an embedding, from Boltzmann sampling at two temperatures (see legend). The embedding is of a fully-connected graph, topology as described in Sect.~\ref{sect:embedding_graph}, where each logical spin is made up of 3 spins ($K=3$) and $J_F=-2.0$. Couplings $J_{ij}$ and local fields $h_i$ chosen uniformly randomly from $[-1:0.2:1]$\footnote{-1 to +1 in steps of 0.2.}.
    The solid lines represent the expected decay in $P_L$ from theory (Eq.~(\ref{eq:prob_logical})). We see a clear exponential decay with problem size $N$. 
    Each data point is averaged over 100 random instances. Error bars are one standard deviation over the problem instances. For each instance, we compute the exact $P_L$ by iterating over all configurations of the embedded problem (for $N=10$ the embedded problem contains $2^{30}$ configurations).
    }
    \label{fig:prob_logical}
\end{figure}

We see this striking unfavorable exponential scaling in Fig.~\ref{fig:prob_logical}, which for a fully connected graph under the embedding parameters and temperatures we study, demonstrates once the size is above around $N\approx 120$, only around one sample per billion will be from the logical subspace. It is therefore prohibitive to simply discard solutions from outside of the logical subspace, for any problem of even modest size (e.g. 100 spins). 
The hotter the distribution, the worse the scaling and the more likely it is to leave the logical subspace. We study the exponential scaling in the next subsection.

Moreover, hardware or algorithmic constraints on 1) the logical spin strengths, $J_F$,  2) the size of the logical spins $K$, by way of the connectivity of the embedded graph, and 3) the temperature, also impose difficulties in skirting around this problem by using favorable parameter setting choices for the embedding or picking a low enough temperature.

Whilst the origin of points 2) and 3) is clear (fixed hardware graph and cooling limitations), we elaborate on 1). There are two factors to consider here:
\begin{itemize}
    \item[i)] Physical device constraints may determine the maximum absolute value of any coupling, in particular $|J_F|$, as this is related to the maximum physical energy scale of the system, and can not be increased arbitrarily. On the current generation of D-Wave quantum annealing device all couplings must have values within $[-2,+1]$ (in units of the fixed system energy scale).
    Nominally this limits $|J_F|$ to twice the size of problem couplings (typically chosen in the full range $[-1,1]$). 
    One can achieve relative values of $|J_F|/J_{MAX}>2$, where $J_{MAX}$ is the maximum value of the logical problem couplings, by reducing the problem variables' magnitude by a constant factor. Since the temperature is fixed, this effectively increases the sampling temperature within the logical subspace, i.e. the distribution tends to the trivial uniform distribution as the ratio is increased. 
    In principle this problem can be solved if the temperature of the sampler can be freely tuned, but clearly this is not feasible in physical devices.
    This, in addition to point ii) below leads to the notion of an optimal (non-infinite) $J_F/J_{MAX}$ ratio for optimization purposes \cite{Venturelli:2014nx}. 
    Moreover, since annealing devices are analog in nature and  each coupling is programmed in with fixed precision (normally distributed about the specified value), problem misspecification is increased by reducing the problem scale, and can cause a dramatic reduction in sampling quality \cite{analog_errors,analog_errors2}. 
    \item[ii)] As mentioned previously, the landscape can become challenging to traverse if $|J_F|$ is too large. Since thermal algorithms, including non-zero temperature quantum annealing, must traverse the landscape of all possible solutions, introducing large energy penalties (barriers) between different configurations can cause the system to become stuck in suboptimal regions, failing to thermalize properly. Though one may hope that quantum tunneling will help if the barriers in the effective potential are thin enough \cite{Albash-scaling}, whether or not this is the case here is not clear, in particular as an anneal approaches the end of the evolution where the transverse field (and thus tunneling rate) is diminishing.
    
\end{itemize}

We will now provide a counting argument which demonstrates these issues more precisely.

\subsection{Analytic expression for relative subspace sampling}\label{sect:analytic}
\begin{figure}
	\centering
	\includegraphics[width=0.95\columnwidth]{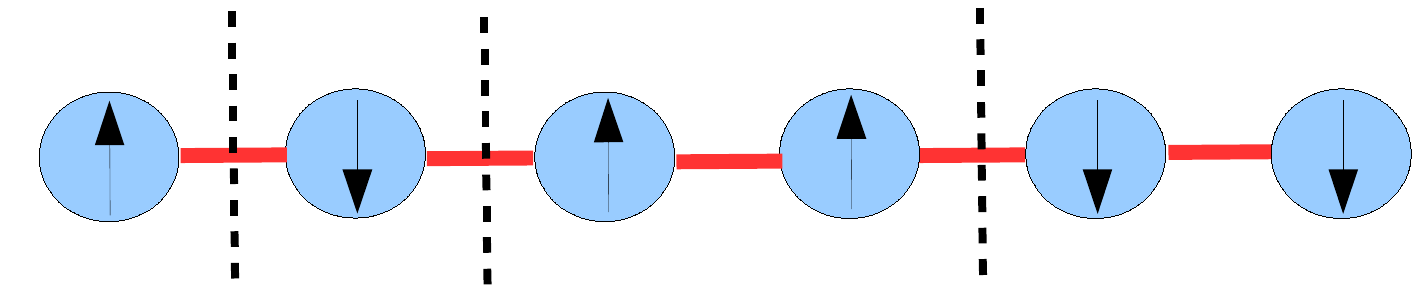}
	\caption{Example of spin configuration of chain of size $K=6$ with $n_{dw}=3$ domain walls. Vertical dash lines represent positions of the domain walls where the spin value changes between sites.
		There can be at most $K-1$ domain walls.
		The red links represent couplings $J_F$.
		The energy increase (penalty) for introducing $n_{dw}$ domain walls is $2n_{dw}|J_F|$. There are $2 {K-1 \choose n_{dw}}$ possible configurations of a spin chain with $n_{dw}$ domain walls.
	}
	\label{fig:domain-wall}
\end{figure}

Let us assume for simplicity that each logical subtree is in fact a path of the same length; i.e. a linear chain, composed of $K$ vertices. We denote by $J_F<0$ the ferromagnetic bonds linking the spins together. 
The native problem size is $N$, and therefore, the embedded version contains $N\times K$ spins (vertices). 
We now estimate the relative sampling weight between subspaces with $n$ broken logical spins (i.e.~chains with not all identically aligned spins), under a Boltzmann distribution at inverse temperature $\beta$. 
In particular, we want to obtain $P_n$, where $P_n$ is the probability of sampling from the subspace with $n$ broken logical spins. 
This quantity will, of course, depend on details of the specific Hamiltonian, that is on the couplings $J_F,\{J_{ij}\},\{h_i\}$ we are considering. To obtain an estimate of that, we consider its average with respect to the values of the couplings $J_{ij}$ and of the local fields $h_i$, assuming that these random variables are independent and identically distributed with a symmetric probability density function. For simplicity, let us assume that their mean is zero. 
Now, consider two configurations, $\sigma^{(\ell)}$ and $\tilde{\sigma}^{(\ell)}$, with $\ell$ domain walls distributed over the chains (i.e. number of positions where the spin flips from one site to the neighbor within the chains). See Fig.~\ref{fig:domain-wall}. Notice that $0 \le \ell \le N (K-1)$. 
Let us relate the spin values of $\tilde{\sigma}^{(\ell)}_i$ and ${\sigma}^{(\ell)}_i = \xi_i \tilde{\sigma}^{(\ell)}_i$ by the vector $\xi$, where $\xi_i$ is $+1$ if ${\sigma}^{(\ell)}_i = \tilde{\sigma}^{(\ell)}_i$, and else, $-1$ (where $i=1,\dots, NK$).
We have, labeling with $p(\sigma)$ the probability averaged over the values of the couplings $J_{ij}$ and $h_i$ (``disorder'') of the configuration $\sigma$,
\begin{equation}\label{eq:p_ell}
	p(\sigma^{(\ell)}) = \overline{\frac{e^{-\beta H(\sigma^{(\ell)})}}{Z}}
\end{equation}
where $Z=\sum_\sigma \exp(-\beta H(\sigma))$ is the partition function,
and the overline denotes the average over the disorder.
By re-defining couplings via $J_{ij}\sigma_i \sigma_j = \tilde{J}_{ij}\tilde{\sigma}_i \tilde{\sigma}_j$, where $\tilde{J}_{ij}=\xi_i \xi_j J_{ij}$ (similar for $J_F$ and $h_i$), we can relate $p(\sigma^{(\ell)})$ and $p(\tilde{\sigma}^{(\ell)})$.
In particular, as shown explicitly in Appendix \ref{sect:appendix_probability}, we have
\begin{equation}\label{eq:prob_manychains}
	p(\tilde{\sigma}^{(\ell)}) = \overline{\frac{e^{-\beta H(\tilde{\sigma}^{(\ell)})}}{Z}} = \overline{\frac{e^{-\beta H(\sigma^{(\ell)})}}{Z'}}
\end{equation}
where $Z' = Z'(\beta,H,\xi)$ differs from $Z=Z(\beta,H)$ through the re-mapping of variables via $\xi$.
This calculation uses the fact that the average over the disorder is done with a probability density function which is symmetric with respect to a sign flip of each coupling $J_{ij}$ and $h_{i}$ (see Appendix \ref{sect:appendix_probability}).
Unfortunately, the change of sign of some of the couplings has the effect of changing the partition function $Z\rightarrow Z'$, and this is due to the fact that the ferromagnetic couplings $J_F$ are fixed and we are not averaging on their value.

\begin{figure}
	\centering
	\includegraphics[width=0.95\columnwidth]{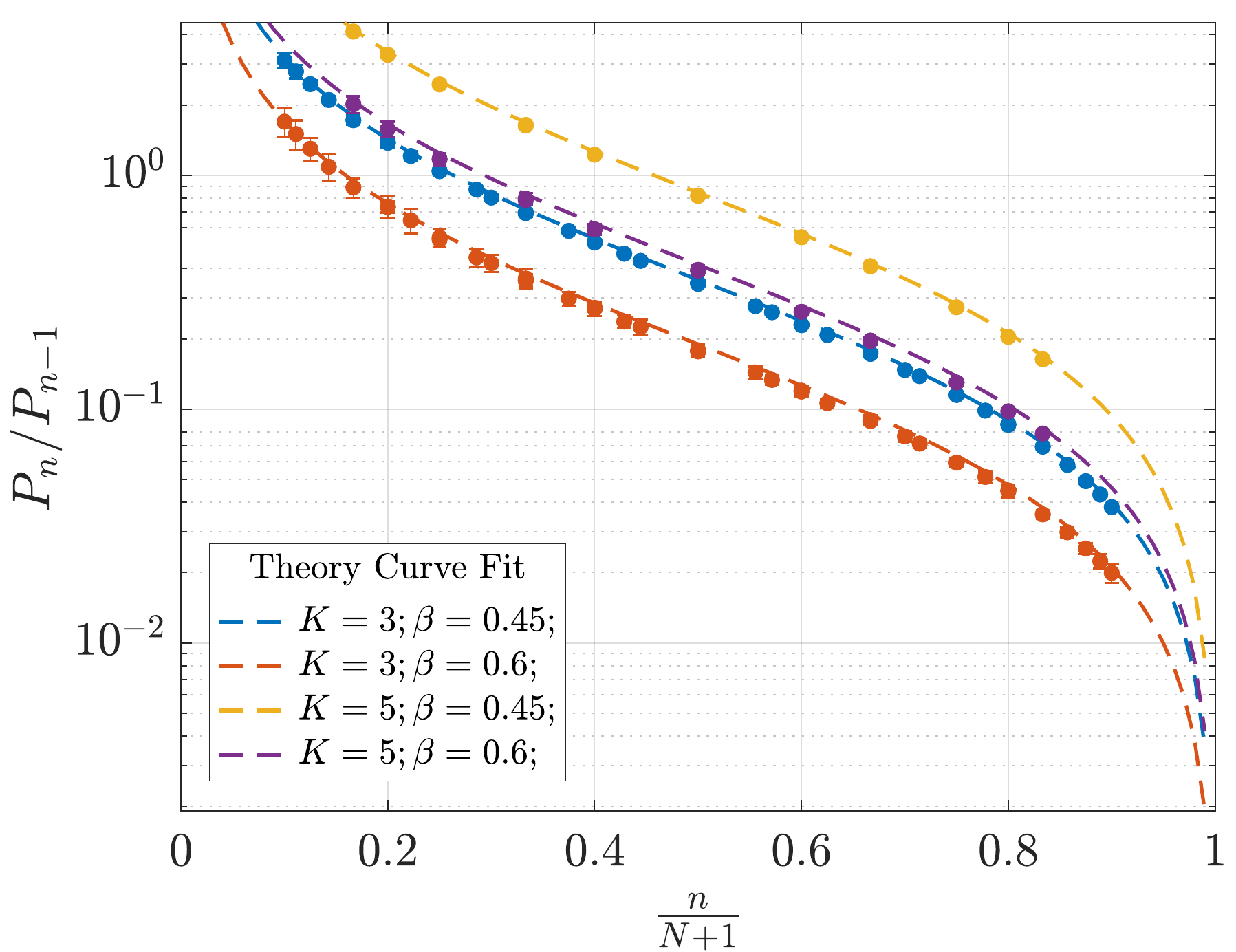}
	\caption{We compare our theoretical Eq.~(\ref{eq:prob_ratio}) (dash lines) to numerical simulations, where $P_n$ is the probability of observing a configuration with $n$ broken logical spins. Each data point is an average over 100 random embedded problems for various choices of $N,n$, and with parameters given in the legend. Error bars are standard deviation. We fix $|J_F|=2$ in units of the original Hamiltonian for all data points. 
	}
	\label{fig:ratio_numBroken}
\end{figure}

To strongly simplify our equations, and ultimately allow us to estimate $P_n$, we consider the so-called \emph{annealed} approximation (see, for example, Ref.~\cite{Castellani2005}), which consists in considering the couplings $J_{ij}$ and $h_{i}$ as dynamical variables, on the same footing of the spin variables. In this case
\begin{equation}
	Z_{\text{ann}} = \overline{Z} = \overline{\sum_{\sigma} e^{-\beta H(\sigma)}},
\end{equation}
and with $Z=Z_{\text{ann}}=Z'$, we obtain
\begin{equation}
	p(\sigma^{(\ell)}) = p(\tilde{\sigma}^{(\ell)}).
\end{equation}
Therefore, under the annealed approximation, the probability of a configuration (averaged over the disorder) depends only on the number of domain walls. 
If we call $p_\ell$ the probability of a configuration with $\ell$ domain walls, we have
\begin{equation}
	p_\ell = e^{2 \beta \ell J_F} p_0.
\end{equation}
This fact, together with the fact that there are
\begin{equation}
	\mathcal{N}(\ell)  = 2^N \binom{(K-1) N}{\ell},
\end{equation}
possible configurations with $\ell$ domain walls, allow us to write for the total probability of observing $\ell$ domain walls $P^{(\ell)}$:
\begin{equation}
	P^{(\ell)} = \mathcal{N}(\ell) \, p_\ell = \binom{(K-1) N}{\ell} e^{2 \beta \ell J_F } P_0,
\end{equation}
where $P_0$ is the probability to sample a configuration from the logical subspace (summed over all configurations and averaged over the disorder). In other words, $P_0 = 2^N p_0$ since there are $2^N$ possible logical configurations. For the probability to observe a state outside the logical subspace $P_{\text{out}}$, we have, by the binomial theorem,
\begin{equation}
	P_{\text{out}} = \sum_{\ell=1}^{(K-1) N} P^{(\ell)} = P_0 \left( (1 + e^{2 \beta J_F})^{(K-1)N} - 1 \right).
\end{equation}
Therefore, using that $P_0+P_{\text{out}} = 1$,
\begin{equation}\label{eq:prob_logical}
	P_0 = \left(1 + e^{2 \beta J_F} \right)^{-(K-1) N}.
\end{equation}
Let us now turn to the general case, that is the computation of the probability of observing $n$ broken chains. We have
\begin{equation}\label{eq:prob_broken}
\begin{split}
	P_n & = \binom{N}{n} \sum_{q_1 = 1}^{K-1} \cdots \sum_{q_n = 1}^{K-1} 2^N  \binom{K-1}{q_1} \cdots \binom{K-1}{q_n} \times \\
	& \qquad \times p_{q_1 + \dots + q_n},
\end{split}
\end{equation}
where the first binomial coefficient comes from the choice of $n$ chains to break (among $N$ available), the term $2^N  \binom{K-1}{q_1} \cdots \binom{K-1}{q_n}$ are the possible configurations of $n$ chains with $q_1, \dots, q_n$ domain walls respectively, and $p_{q_1+\dots+q_n}$ is the probability of observing $q_1+\dots+q_n$ domain walls. We obtain
\begin{equation}
\begin{split}
\label{eq:prob_broken2}
	P_n & = \binom{N}{n} \sum_{q_1 = 1}^{K-1} \cdots \sum_{q_n = 1}^{K-1} 2^N  \binom{K-1}{q_1} \cdots \binom{K-1}{q_n} \times \\
	& \qquad \times e^{2 \beta J_F (q_1+\cdots+q_n) } p_0\\
	& = \binom{N}{n} P_0 \left( \sum_{q = 1}^{K-1} \binom{K-1}{q} e^{2 \beta J_F q} \right)^n\\
	& = \binom{N}{n} \frac{\left( \mathcal{P}_w \right)^n}{\left(\mathcal{P}_w + 1 \right)^{N}},
\end{split}
\end{equation}
where
\begin{equation}
\label{eq:penalty_weight}
	\mathcal{P}_w = (1 + e^{2 \beta J_F})^{K-1} - 1.
\end{equation}
In particular, 
\begin{equation} \label{eq:prob_ratio}
	\frac{P_n}{P_{n-1}} = \left( \frac{N+1}{n}-1\right) \mathcal{P}_w,
\end{equation}
and we demonstrate the success of this equation, and so of the annealed approximation for our case, in Fig.~\ref{fig:ratio_numBroken}, plotting for several parameter choices $P_n/P_{n-1}$ as a function of $n/(N+1)$.

We now make some brief comments on these relations:
\begin{itemize}
	\item [i)] Eqs.~(\ref{eq:prob_logical}) and (\ref{eq:prob_broken2}) are trivially exact for $\beta \rightarrow 0$, since in this case all configurations are sampled equally. In general, the annealed approximation is correct in the thermodynamical limit as long as the partition function is a self-averaging quantity. This happens above the critical temperature of the spin glass transition.
	\item [ii)] One consequence of our assumptions is that $|J_F|$ must be large enough so the global-to-global property holds, i.e. $P_0\rightarrow 1$ as $\beta \rightarrow \infty$. In particular, if $|J_F|\to\infty$ then Eq.~(\ref{eq:prob_logical}) is correct since $P_0\to1$, and on the other hand if $|J_F|=0$ again Eq.~(\ref{eq:prob_logical}) gives the correct result, that is each configuration has the same probability and therefore $P_0=2^N/2^{NK}$. The same, correct result is obtained for $\beta=0$, where the annealed approximation is known to be exact. However, in general it is unclear the extent to which the annealed approximation gives an incorrect result in our computation for arbitrary temperatures or problem sizes (also see Sect.~\ref{sect:discussion} for more discussion).
	Lastly, it is clear that if the global-to-global property does not hold, Eqs.~(\ref{eq:prob_logical}), (\ref{eq:prob_broken2}) will not be valid at low enough temperatures.
	\item [iii)] The probability $P_0$ decays exponentially in problem size, and chain size. Thus there can be huge sampling benefits from utilizing more efficient embeddings with smaller chains.
	Compatible with intuition we see logical subspace sampling can be improved for larger $\beta |J_F|$ (colder temperature and/or stronger ferromagnetic couplings).
	\item [iv)] For hardware constrained $\beta$ and $J_F$ (i.e.  can not scale with $N$), it is clear that for large enough problems, and ones with more complicated embeddings (larger $K$), there will inevitably be troubles sampling the logical subspace directly.
	In Fig~\ref{fig:prob_logical} we show the decay of $P_0$ as a function of $N$, with $K=3$, for two temperatures. The theory of Eq.~(\ref{eq:prob_logical}) matches rather well with the numerical data, giving us confidence about the assumptions we made in our derivation, for the chosen parameters.
\end{itemize}

In the next subsection, we demonstrate the difficulty of solving this problem through a simple, but tractable, model.

\subsection{Projection techniques and sampling bias}\label{sect:bias}

In this subsection, we describe limitations on postprocessing techniques that project from the embedded space back to the logical space. Specifically, we demonstrate by example that under reasonable assumptions on such projections, sampling bias is unavoidable. The example is simple and not contrived, suggesting that this bias is generally hard to avoid. The assumptions we make on the postprocessing are that 1) the temperature of the Boltzmann distribution we are aiming for remains the same as for the logical subspace, 2) ``if it ain't broke, don't fix it" -- we do not adjust the values of any spins from non-broken logical spins, 3) we do not discard solutions, and 4) we carry out the projection one solution at a time. These assumptions are motivated by the need to keep the postprocessing computational effort tractable and to avoid trivial solutions to the problem, such as providing Boltzmann samples at infinite temperature. It might be interesting to see if relaxing some of them, while keeping the computational effort reasonable, can lead to less bias or if one can prove that relaxing the assumptions does not help. These assumptions already encompass the leading postprocessing approach, majority vote, and allow for significantly broader approaches. In the next section, we will numerically demonstrate the significant bias resulting from majority voting, and provide an alternative that does better. 

We prove the impossibility of postprocessing without biasing the sampling, under the assumptions above, by showing its impossibility for a simple case, i.e. through a counter example. Consider an $N$ spin problem which is embedded by replacing one of its nodes with two nodes, resulting in an $N+1$ spin problem. The postprocessing task is to provide means to decide, given
a configuration in which the two spins in the logical spin do not align, with what probability they should be projected to both spin up, or both spin down (fixing the value of all other spins). The hope would be that after this projection, and with sufficiently many samples, the distribution is still Boltzmann at the same temperature. 

Let us call $C$ the configuration of the fixed $N-1$ spins, and $C_{-1,1}, C_{1,-1}, C_{1,1}, C_{-1,-1}$ the full configuration of $N+1$ fixing the $N-1$ spins as in $C$, with the subscript denoting the configuration of the logical spin.
Similarly, we call the cost of these configurations $E_{a,b}^{(C)}$ for $a,b \in \{-1,1\}$. With probability $P_{(a,b)\rightarrow (c,c)}^{(C)}$ configuration $C_{a,b}$ is projected to $C_{c,c}$ ($c \in \{-1,1\}$). If the logical spin is aligned, we should not change it; $P_{(a,a)\rightarrow (a,a)}^{(C)}=1$.
The probability to observe configuration $C_{a,b}$, before any projection, is $\exp(-\beta E_{a,b}^{(C)} ) /Z$ where 
\begin{equation}
    Z = \sum_c \sum_{a,b=\pm 1 }e^{-\beta E^{(c)}_{a,b}}
\end{equation}
is the partition function for normalization.

Let us assume there does exist a procedure to re-map the probabilities such that they still follow a Boltzmann distribution at the same temperature. Then we have:

\begin{widetext}
\begin{equation}
\begin{split}
\label{eq:mapping_probs}
& P_{-1,-1}^{(C)} = \frac{1}{Z}\left[ e^{-\beta E_{-1,-1}^{(C)}} + P_{(-1,1)\rightarrow (-1,-1)}^{(C)}\, e^{-\beta E_{-1,1}^{(C)}} +P_{(1,-1)\rightarrow (-1,-1)}^{(C)}\, e^{-\beta E_{1,-1}^{(C)}} \right] \stackrel{!}{=} \frac{e^{-{\beta} E_{-1,-1}^{(C)}}}{Z_L} \\
& P_{1,1}^{(C)}= \frac{1}{Z}\left[ e^{-\beta E_{1,1}^{(C)}} + P_{(-1,1)\rightarrow (1,1)}^{(C)}\, e^{-\beta E_{-1,1}^{(C)}} +P_{(1,-1)\rightarrow (1,1)}^{(C)}\, e^{-\beta E_{1,-1}^{(C)}} \right] \stackrel{!}{=} \frac{e^{-{\beta} E_{1,1}^{(C)}}}{Z_L} .
\end{split}
\end{equation}
\end{widetext}
The second equals sign is used to indicate we require that $C_{-1,-1}, C_{1,1}$ are sampled from a Boltzmann distribution with corresponding partition function over the logical subspace
\begin{equation}
    Z_L = \sum_c \sum_{a=\pm 1} e^{-\beta E^{(c)}_{a,a}}.
\end{equation}
For now, let us assume no solutions are discarded, so that $P_{(-1,1)\rightarrow (-1,-1)}^{(C)} + P_{(-1,1)\rightarrow (1,1)}^{(C)} = 1$ (and similar for $C_{(1,-1)}$). In this case, these two equations, with two unknowns, can be solved.

One finds
\begin{widetext}
\begin{equation}
\label{eq:linear-sol}
\begin{split}
& P_{(1,-1) \rightarrow (-1,-1)}^{(C)} =  - P_{(-1,1)\rightarrow (-1,-1)}^{(C)}e^{-\beta (E_{-1,1}^{(C)}-E_{1,-1}^{(C)})} + \left[  \frac{Z}{Z_L}e^{-{\beta} E_{-1,-1}^{(C)}}  - e^{-\beta E_{-1,-1}^{(C)}} \right]e^{\beta E_{1,-1}^{(C)}} \\ 
& P_{(1,-1) \rightarrow (-1,-1)}^{(C)} = - P_{(-1,1)\rightarrow (-1,-1)}^{(C)}e^{-\beta (E_{-1,1}^{(C)}-E_{1,-1}^{(C)})}  + \left[-\frac{Z}{Z_L}e^{-{\beta} E_{1,1}^{(C)}}  + e^{-\beta E_{-1,1}^{(C)}} + e^{-\beta E_{1,1}^{(C)}} + e^{-\beta E_{1,-1}^{(C)}}\right]e^{\beta E_{1,-1}^{(C)}}
\end{split}
\end{equation}
\end{widetext}

which specifies two linear equations with the same gradients, but, in general, different intercept values, which therefore have no solutions. 
To see this, compare the ratio $Z/Z_L$ from solving Eqs.~(\ref{eq:linear-sol}),
\begin{equation}
    \frac{Z}{Z_L} = 1 + \frac{e^{-\beta E_{-1,1}^{(C)}}+ e^{-\beta E_{1,-1}^{(C)}}}{e^{-\beta E_{-1,-1}^{(C)}}+e^{-\beta E_{1,1}^{(C)}}},
    \label{eq:ratio1}
\end{equation}
with the exact
\begin{equation}
    \frac{Z}{Z_L} = 1 + \frac{\sum_{c} e^{-\beta E_{-1,1}^{(c)}}+ e^{-\beta E_{1,-1}^{(c)}}}{\sum_{c} e^{-\beta E_{-1,-1}^{(c)}} + e^{-\beta E_{1,1}^{(c)}}},
    \label{eq:ratio2}
\end{equation}
which depends on all possible configurations $c$, and not just the single configuration $C$. In general, Eqs.~(\ref{eq:ratio1}) and (\ref{eq:ratio2}) will not be the same, meaning the Eqs.~(\ref{eq:mapping_probs}) cannot be simultaneously satisfied. We demonstrate this by example.

We show that even in the simplest case, in which the Hamiltonian gives a ferromagnetic ring on $N$ spins, with $N$ odd, that Eqs.~(\ref{eq:ratio1}), (\ref{eq:ratio2}) are violated.
The embedded Hamiltonian on $N+1$ spins is then
\begin{equation}
    H = -|J_F|s_0s_1 - \sum_{i=1}^N s_is_{i+1}
\end{equation}
where we identify $s_{N+1}\equiv s_0$. There are $N+1$ total spins. 
Let $C_i$ denote a configuration of the $N-1$ spins labelled $2,\dots N$.
We take $C_1=(-1,\dots ,-1)$, and  $C_2=(-1,+1,-1,+1,\dots,-1,+1)$ (assume $N$ is odd).

We compute the energies $E^{(C_{1,2})}_{(\pm1, \pm 1)}$, where the subscript is the spin value for $(s_0,s_1)$, in Table \ref{table:energies}.

\begin{widetext}
\begin{center}
\begin{table}[h]
\begin{tabular}{ |c | c c c c| } 
 \hline
   & $(-1,-1)$ & (-1,+1) & (+1,-1) & (+1,+1) \\ 
  \hline
 $C_1$ & $-N-|J_F|$ & $-N+2+|J_F|$ & $-N+2+|J_F|$ & $-N+4-|J_F|$ \\ 
 $C_2$ & $N-2-|J_F|$ & $N-4+|J_F|$ & $N+|J_F|$ & $N-2-|J_F|$  \\ 
 \hline
\end{tabular}
\caption{Table of the energies $E_{(\pm 1, \pm 1)}^{C_{1,2}}$.}
\label{table:energies}
\end{table}
\end{center}
\end{widetext}

Now consider the quantity $r(C):=\frac{Z}{Z_L}-1$ computed using the configurations $C_1$ and $C_2$ from Eq.~(\ref{eq:ratio1}):
\begin{equation}
\begin{split}
    & r(C_1) = \frac{e^{-2\beta |J_F|}}{\cosh{2\beta}} \\
    & r(C_2) = e^{-2\beta |J_F|}\cosh{2\beta}.
\end{split}
\end{equation}
We have $r(C_1)\neq r(C_2)$ (except for the very particular case $\beta = 0$), while the quantity $\frac{Z}{Z_L}-1$ has to be configuration-independent as we can see from Eq.~\eqref{eq:ratio2}.

Interestingly, in this case even knowing $Z$ and $Z_L$ is not enough to solve this problem. Of course this does not exclude the possibility of obtaining Boltzmann samples from an embedded distribution by relaxing at lease one of
the restrictions we imposed: 
1) one may not require the final distribution is at the same temperature of the sampler, 2) one could use additional information about the structure of the problem, 3) one can discard certain configurations, or 4) performing post-processing on a large set of configurations.

Whilst the above argument indicates it is difficult, or impossible, to perfectly recover the target distribution, it is not clear the extent to which sampling can be biased by certain projection techniques. 
In the next sections we numerically study some examples.

\section{{Post-processing techniques and numerical} results}

\subsection{Majority voting}\label{sect:MV}
In the context of optimization tasks, one will often use majority vote (MV) to obtain relevant solutions when \textit{illogical} configurations (configurations outside of the logical subspace) are present in the sampling. This procedure is easy to implement and understand. Given a single configuration, for each logical spin which is not aligned identically, correct it by going with the majority. If there is a tie, one can pick at random. For optimization purposes, this is a simple way to obtain a greater number of solutions and does not cause any intrinsic issues. For sampling however, this introduces biases in the sampling rate of certain logical configurations.

We first demonstrate this by example using an embedding of a fully connected graph, where each variable becomes a logical spin of size $K$ (see Fig.~\ref{fig:fc_cartoon}).
The problems we study have values $J_{ij}$ and $h_i$ chosen uniformly randomly from $[-1:0.2:1]$ ($-1$ to $+1$ with step size 0.2).
We restrict our analysis for now to small sizes so we can exactly compute the probabilities of each configuration (i.e. compute the partition function). As a result, the largest system we study is $8 \times 3 = 24$ variables.
In order to demonstrate the sampling bias for these small (numerically exactly solvable) problems, we take the temperature parameter $\beta = 0.6$. In general, colder temperatures will exhibit less bias (assuming the global-to-global property), by the arguments of the previous section.

\begin{figure}
    \centering
    \includegraphics[width=0.98\columnwidth]{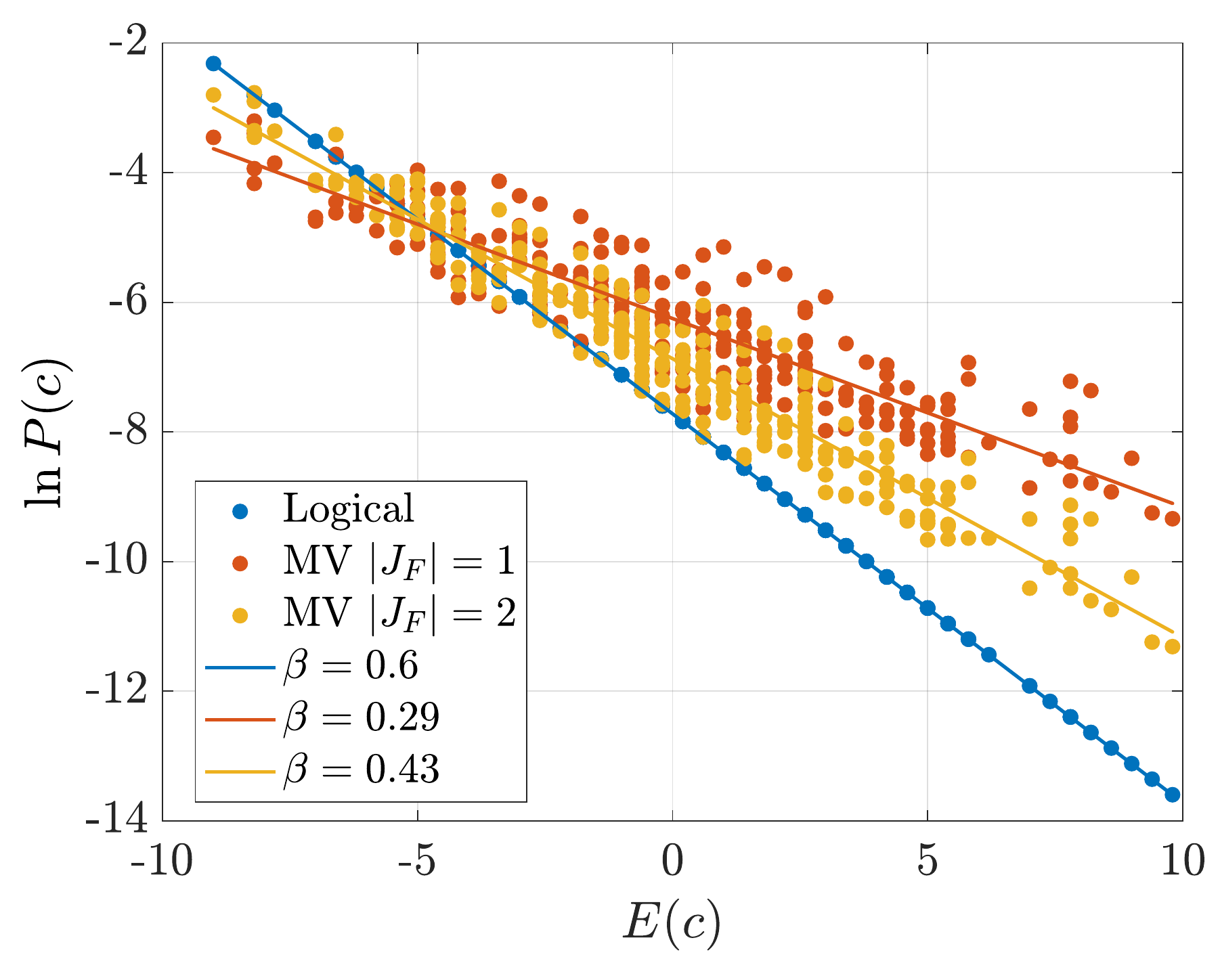}
    \caption{The effect of majority vote (MV) on sampling for an 8 variable fully connected problem. Here, $E(c)$ is the cost associated with logical configuration $c$, and $P(c)$ the corresponding sampling probability under a Boltzmann distribution. In the embedding, each variable becomes a logical variable of size $K=3$ (see Fig.~\ref{fig:fc_cartoon}). We demonstrate with two different ferromagnetic coupling strengths, in units of the native Hamiltonian $H$. The straight lines are found by least squares fitting, where the gradient represents the inverse temperature (see legend).}
    \label{fig:bman_mv}
\end{figure}

Our analysis shows that in general, and unsurprisingly, performing majority vote induces biases into the sampling procedure, even when the ferromagnetic couplings are `strong' (e.g. twice the magnitude of any coupling in the underlying Hamiltonian, as is the case in typical implementations on current hardware, such as the D-Wave 2000Q). 
An example of this is shown in Fig.~\ref{fig:bman_mv} where one can notice a few distinctive features. 1) The distribution after performing MV is not a Boltzmann distribution as the points do not lie on a straight line.
2) Moreover, there exist configurations of the same cost, but different sampling rates. 3) Assigning the best fit temperature to the distribution gives a hotter distribution compared to the sampling temperature; in particular, it tends to flatten out the distribution.

Indeed, in light of the discussion in Sect.~\ref{sect:problem_embedding} it is not surprising MV fails as it comes under a special case of the argument outlined which shows it is not possible in general to perform such a mapping. What is perhaps not obvious is how poorly MV can perform, failing to capture much semblance of a Boltzmann distribution at all by biasing the statistics. We restricted ourselves to small sizes so that we could perform the computations exactly (i.e. analyzing all $2^{N K}$ configurations), but our analysis also indicates that in general the biases associated with MV become more detrimental with size. 

\begin{figure}
    \centering
    \includegraphics[width=0.98\columnwidth]{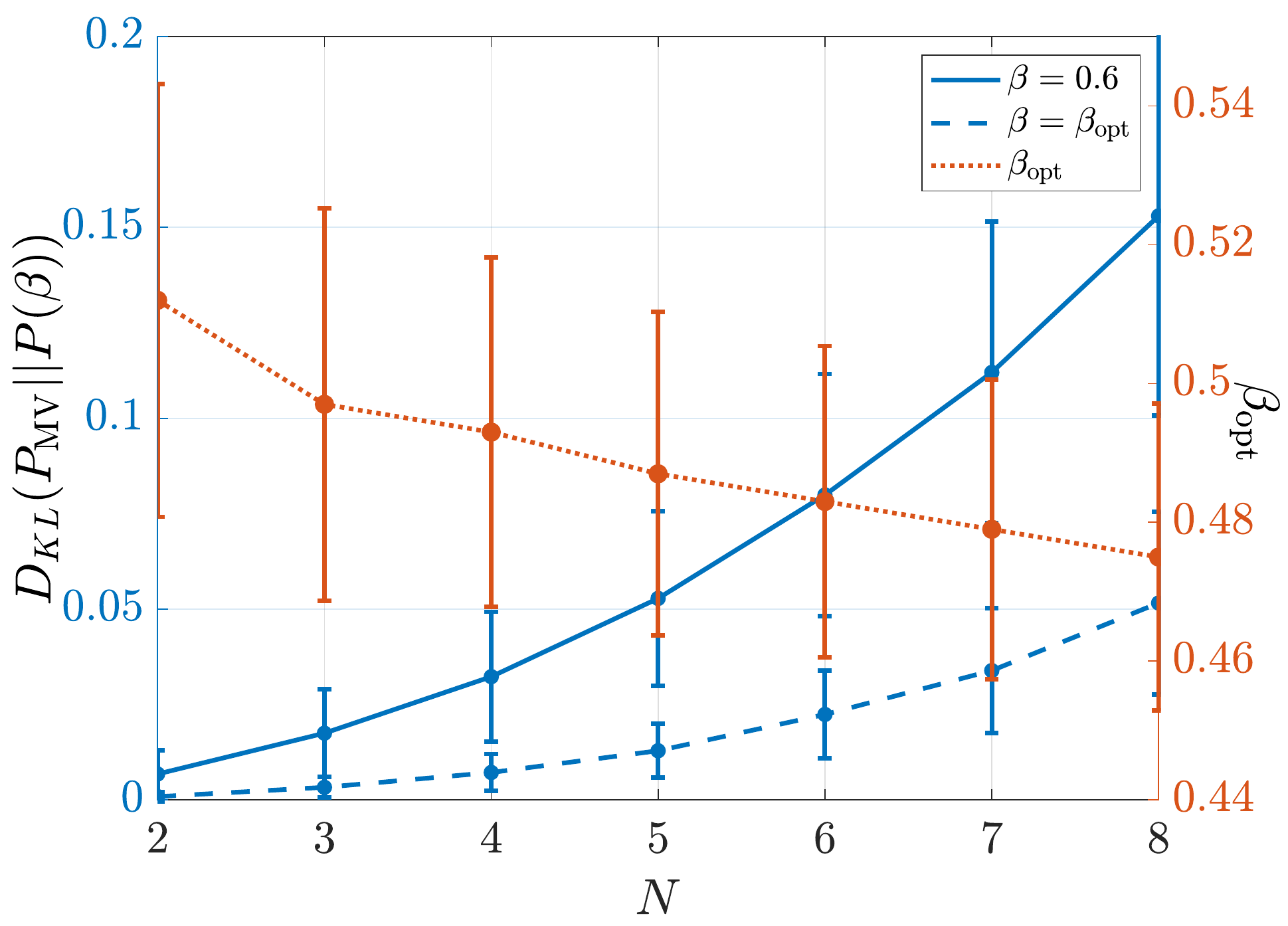}
    \caption{KL divergence of majority voted distribution to Boltzmann distribution $P(\beta)$ at inverse temperature $\beta$, as a function of problem size $N$ (number of variables in fully connected graph). The sampling of the embedded problem was performed with $\beta=0.6$. We use logical spins of size $K=3$ for the embedding (as in Fig.~\ref{fig:fc_cartoon}).
    Each data point is averaged over 500 random problems and embeddings.
    The solid blue curve is the KL divergence between the MV data and the `ideal' Boltzmann distribution (i.e. if no embedding was required).
    The dash blue curve is the KL divergence between the MV data and a Boltzmann distribution at the optimal inverse temperature $\beta_{\mathrm{opt}}$ (which is found, for each problem, by minimizing the KL divergence).
    The dotted red line (right y-axis) is the optimal fitting inverse temperature.  Error bars are standard deviation. Here $|J_F|=2$ in units of the Hamiltonian.}
    \label{fig:mv_scaling}
\end{figure}

In Fig.~\ref{fig:mv_scaling} we notice two related effects. Firstly, larger problems are more adversely affected by MV as determined by the KL-divergence at the optimal temperature, and second, this optimal sampling temperature becomes hotter for larger problem sizes. The latter indicates the distribution is becoming flatter as problem size increases.
This is not unexpected since here the temperature and ferromagnetic couplings $J_F$ are not scaling with problem size, and by the arguments in the previous section one therefore expects to observe a greater number of states outside of the logical subspace.

\subsection{A better approach: restricted resampling}
Here we outline a new approach called restricted resampling (RRS) to overcome some of the issues outlined above, inspired by thermal sampling algorithms. 
As before, we assume one receives perfect thermal (Boltzmann) samples of the embedded problem, at some inverse temperature $\beta$ \footnote{One may need to first estimate $\beta$, as discussed in Refs.~\cite{perdomo,global_warming}, or through density of states estimation \cite{wang-landau1,wang-landau2,dos_estimation}.}.
In RRS, one performs a thermal resampling at the designated temperature over a restricted number of problem variables. In particular, when one observes a configuration with $N_B$ broken logical spins, one implements a `resampling' of these variables within the logical space at inverse temperature $\beta$; that is, one effectively performs a Monte Carlo algorithm over a space of size $2^{N_B}$.
Though this does not guarantee to perfectly recover a Boltzmann distribution (again, this algorithm also falls under the arguments outlined in Sect.~\ref{sect:problem_embedding}), we show numerically it clearly outperforms MV.
We therefore propose RRS as an alternative to majority vote and other similar projection techniques.

We outline the general idea of RRS in Algs.~\ref{alg:rrs} and \ref{alg:bmanSubset}. This pseudocode is intended to just give the basic outline of how one could implement RRS, and we stress that any algorithm which can provide thermal samples can be used as the subroutine Alg.~\ref{alg:bmanSubset}. For example, one could use cluster flips instead of single spin flips, or replica-exchange Monte Carlo (parallel tempering), to generate the samples.

In Alg.~\ref{alg:rrs} we first construct the set $B$ of broken logical spins, and also a configuration which respects the spin-values for the logical spins which are not broken. We then thermally resample this configuration at temperature $\beta$, but only resampling over the set of spins $B$.

\begin{algorithm}[H]
\caption{Outline of RRS algorithm. 
The input is a configuration $\tilde{C} \in \tilde{\mathcal{H}}$ from the embedded space, the native Hamiltonian $H$ (over $N$ spin variables), and the desired sampling inverse temperature $\beta$.
In line 5, $V(T_k)$ corresponds to the vertices of the $k$-th logical subtree $T_k$. $S_k$ is therefore the configuration of the $k$-th logical spin. An example implementation of the subroutine BoltzmannSampleOverSubset is given in Alg.~\ref{alg:bmanSubset}.}
\label{alg:rrs}
\begin{algorithmic}[1]
\Procedure{RRS}{$\tilde{C}$,$H$,$\beta$}
\State $C\leftarrow $ Random configuration of length $N$
\State $B$ = [\,] \Comment{Set of broken logical spins}
\For{$k$ = 1 \textbf{to} $N$ }
\State $S_k = \{\tilde{C}_i : i \in V(T_k) \}$
\If{$s_i  = s \, \forall s_i \in S_k$}
\State $C_k = s$
\Else
\State Add $k$ to set $B$
\EndIf
\EndFor

\State \textbf{return} BoltzmannSampleOverSubset($H$,$\beta$,$C$,$B$)
\EndProcedure
\end{algorithmic}
\end{algorithm}

\begin{algorithm}[H]
\caption{Example of implementation for subroutine used in Alg.~\ref{alg:rrs}. In line 4, FlipRandomSpinFromSet($C$,$B$) will flip a spin in configuration $C$, chosen randomly from set $B$. We do not specify explicitly the break condition for the while loop since this is up to user implementation (e.g. after a fixed number of steps, or after the energy landscape has been explored sufficiently).}
\label{alg:bmanSubset}
\begin{algorithmic}[1]
\Procedure{BoltzmannSampleOverSubset}{$H$,$\beta$,$C$,$B$}
\State $E \leftarrow H(C)$ \Comment{Cost (energy) of configuration}
\While{True}
\State $C' \leftarrow$ FlipRandomSpinFromSet($C$,$B$)
\State $E' \leftarrow H(C')$
\If{Random(0,1) $ < \min (1,e^{-\beta (E'-E)})$}
\State $C \leftarrow C'$; $E \leftarrow E'$
\EndIf

\If{break condition True}
\State break
\EndIf
\EndWhile

\State \textbf{return} $C$
\EndProcedure
\end{algorithmic}
\end{algorithm}

\begin{figure}
    \centering
    \includegraphics[width=0.98\columnwidth]{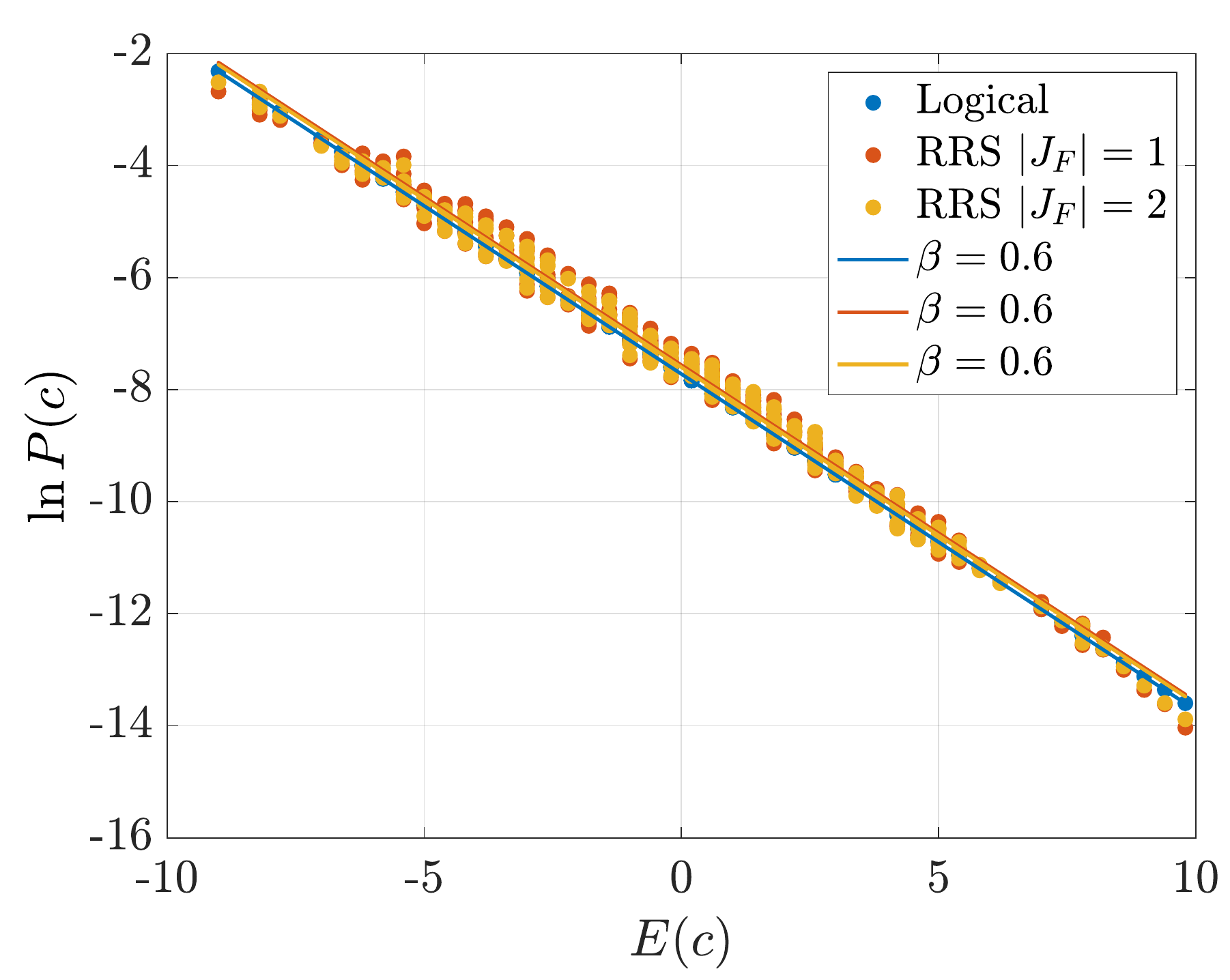}
    \caption{The effect of RRS on sampling for the same 8 variable fully connected problem of Fig.~\ref{fig:bman_mv}. Here, $E(c)$ is the cost associated with logical configuration $c$, and $P(c)$ the corresponding sampling probability under a Boltzmann distribution. In the embedding, each variable becomes a logical variable of size $K=3$ (see Fig.~\ref{fig:fc_cartoon}). The ferromagnetic coupling strengths is units of the native Hamiltonian $H$. The straight lines are found by least squares fitting, where the gradient represents the inverse temperature (see legend). It is clear that RRS outperforms MV.}
    \label{fig:bman_rrs}
\end{figure}

In Fig.~\ref{fig:bman_rrs}, the analogue of Fig.~\ref{fig:bman_mv} of the previous section, we show the effect of RRS for a single problem instance. We see that the resampled distribution is much closer to the ideal as compared to using MV. In particular, the effective temperature after resampling is almost identical to the temperature of the underlying distribution, and configurations of the same cost are sampled with much less variation, as compared to MV.

Note, for our simulations we do this remapping exactly by computing the partition function. In practice, one would need to implement a thermal sampling algorithm, for example based on Monte Carlo techniques.

\begin{figure}
    \centering
    \includegraphics[width=0.98\columnwidth]{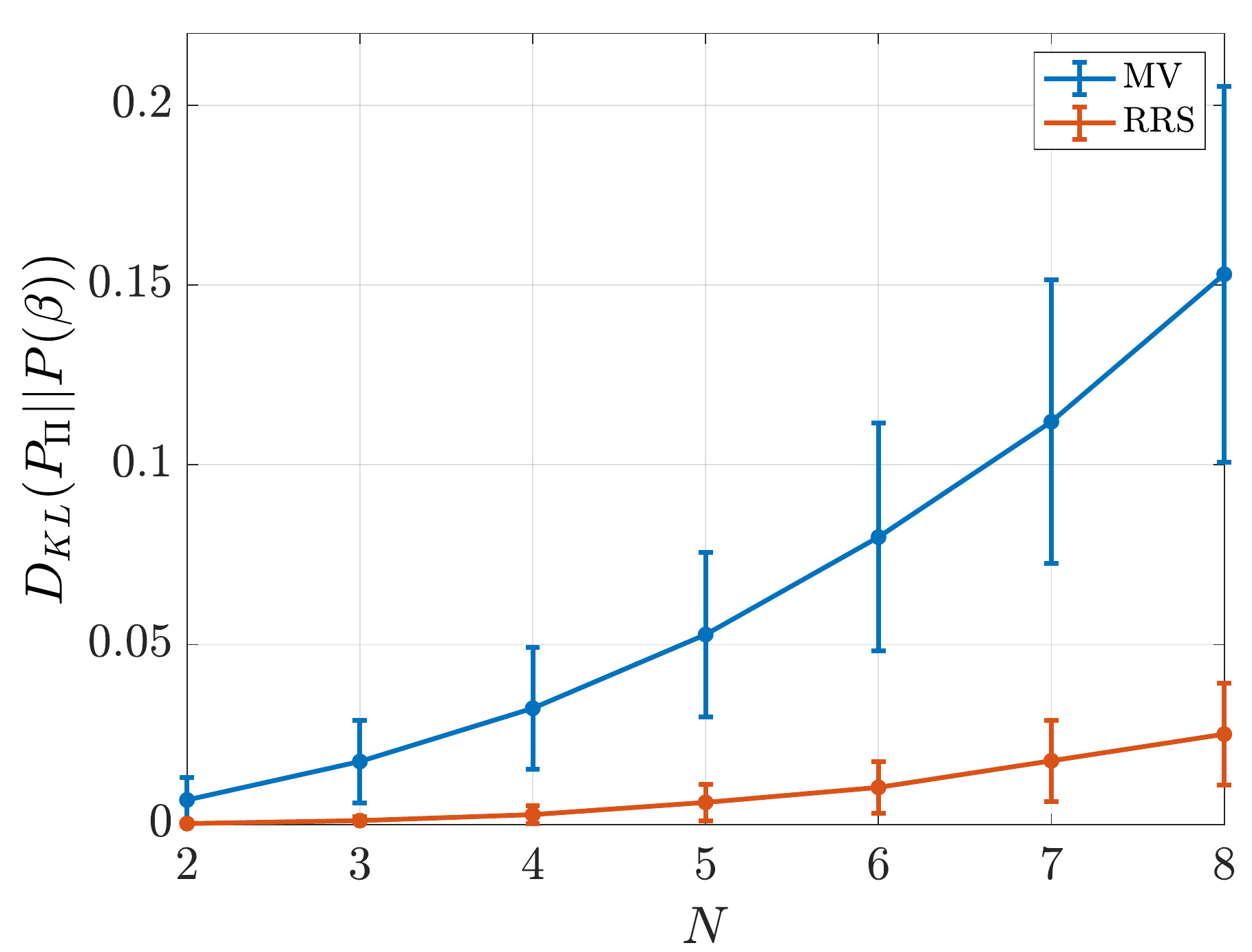}
    \caption{KL divergence to ideal Boltzmann distribution after performing projection $\Pi$ of RRS (red) or MV (blue). $N$ is the native problem size, $\beta=0.6$, with embedding as described in Sect.~\ref{sect:embedding_graph} using $K=3$ and $J_F=-2$. Error bars (standard deviation) are over 500 random samples.}
    \label{fig:rrs_mv}
\end{figure}

\begin{figure}
    \centering
    \includegraphics[width=0.98\columnwidth]{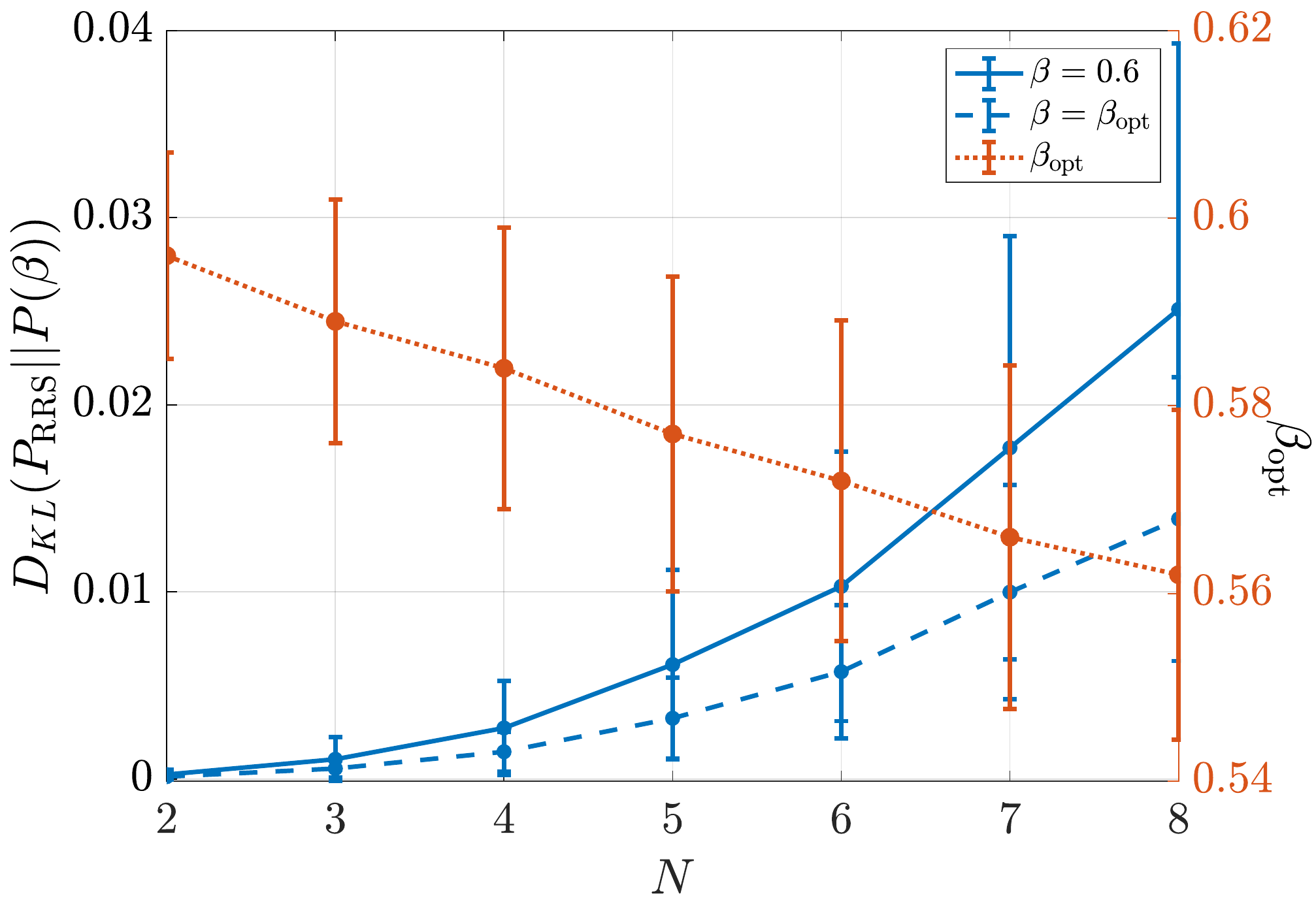}
    \caption{RRS version of Fig.~\ref{fig:mv_scaling}, with the same parameters.
    The effective temperature is much closer to the sampled temperature, although still decreasing with problem size. Similarly, the KL divergence values are less, by around an order of magnitude.
    }
    \label{fig:rrs_scaling}
\end{figure}

In Fig.~\ref{fig:rrs_mv} we see that the scaling of RRS is much more favorable than MV. Moreover, in Fig.~\ref{fig:rrs_scaling} we see the effective sampled temperature after applying RRS is much closer to the physical sampling temperature.

\subsection{Discussion \label{sect:discussion}}

We have identified a potential issue for hardware restricted Boltzmann samplers, such as is proposed for current generation quantum annealers, where embeddings must be used. 
Whilst for strong enough logical spins (ferromagnetic couplings $|J_F|$) and low enough temperatures it is exponentially unlikely in $\beta |J_F|$ to leave the logical space, in reality, these couplings are limited by hardware and do not scale with $N$. In fact, in current hardware such as the D-Wave 2000Q, $|J_F|$ is typically limited to a strength twice that of a problem coupling.
To make matters worse, Ref.~\cite{marshall-rieffel-hen-2017} found that effective sampling temperatures on an experimental quantum annealer tend to increase with problem size.
Embedding therefore inevitably leads to the observation of states which are not in the logical subspace, and since the probability of this occurring nominally scales exponentially in $N$ (Eq.~(\ref{eq:prob_logical})), even for moderately sized systems, one may rarely (or never) observe logical configurations.
These states are not erroneous, caused by errors in the device, but perfectly acceptable configurations in accordance with the Boltzmann distribution of the embedded problem.
The task therefore is, given a sampler which works perfectly, what can be done to project back all configurations to the logical subspace, so that the distribution observed is the desired one (e.g. a Boltzmann distribution).
If these so called illogical states were observed infrequently, a perfectly acceptable solution would be to simply discard these states, since the relative sampling weights are the same in the logical space of the embedded problem, and the native problem (Eq.~(\ref{eq:dist}))

We argued in Sect.~\ref{sect:problem_embedding} that under a reasonable set of assumptions, it is not possible to find such a projection in general which works without error. Our argument assumed that 1) the temperature must remain fixed 2) no illogical configurations are discarded, 3) the projection is performed without knowledge of other configurations, and 4) only broken logical spins are changed.
This includes a wide range of projection algorithms and applies to techniques such as majority vote (MV), and our introduced restricted resampling (RRS) scheme.
This does not preclude the possibility of more advanced schemes where one may violate our assumptions above, for example, collecting many samples first and then performing the projection over the set of samples (e.g. through machine learning techniques), or discarding certain samples.

We have shown that one commonly used technique in the setting of optimization, majority vote, can fail quite spectacularly to capture the intended distribution. The reason for this is it introduces biases to the statistics, and the result is two logical states of the same cost can be sampled at massively different rates (e.g. over an order of magnitude difference in sampling probability). Moreover, the effective temperature after performing MV is much larger than the sampling temperature; i.e. it tends to flatten out the distribution.

We introduced a partial solution to this problem through a scheme called restricted resampling, where one resamples over a restricted set of variables; the ones which are not in the logical space. This not only clearly outperforms MV, but it also gives a distribution with a temperature much closer to the desired one.
This resampling can be performed by a classical algorithm (such as a Monte Carlo type algorithm).

We show another example of a comparison between MV and RRS for a larger problem in Fig.~\ref{fig:sampling_rrs_mv}, where all samples are generated by a Monte Carlo thermal sampler (described in Appendix \ref{sect:appendix_sampler}). This is in contrast to the previous sections where we exactly computed for small sizes the resampling weights for individual configurations. Since estimating the configuration probabilities is infeasible in this case (with $>100$ spins), we focus on estimating the probability of an energy level being sampled $P_i=\frac{g_i}{Z}\exp(-\beta E_i)$.
One can see again that RRS matches closer to the ideal distribution, although there is a large variation between different samples (large error bars), in both cases. Fluctuations in the $P_i$ is due to errors arising from the inexact Monte Carlo implementation, and also due to biases from the projection methods.

\begin{figure}
    \centering
    \includegraphics[width=0.98\columnwidth]{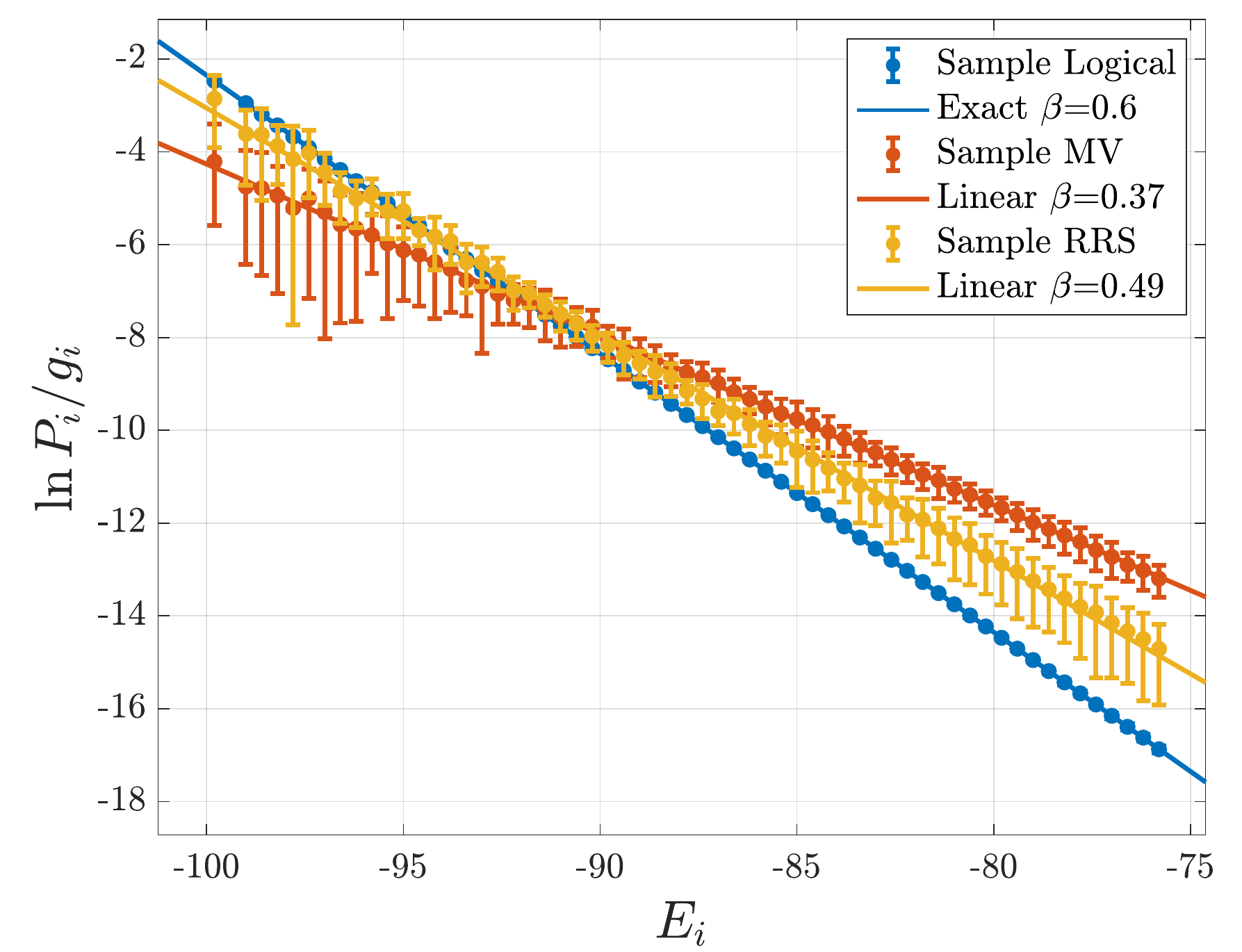}
    \caption{Comparison of RRS and MV for larger problem using Monte Carlo thermal sampler. Here the native problem is fully connected of size $N=35$ with couplings and local fields uniformly random from $[-1:0.2:1]$. Since the native problem is small enough, we can exactly compute the degeneracies $g_i$ for each energy level $E_i$. The blue solid line is the exact profile. $P_i$ is the probability of observing energy level $E_i$ under the sampling.
    The blue dots (with error bars smaller than the dots) is from sampling from the 35 spin problem using a Monte Carlo algorithm with $\beta=0.6$, showing excellent agreement with the exact solid line.
    The red (MV) and yellow (RRS) dots with error bars (standard deviation) are from sampling the embedded problem (topology as in Sect.~\ref{sect:embedding_graph}) with $K=3$ and $J_F=-2$ (in this case the embedded problem contains $35\times 3=105$ spins). 
    The red and yellow solid lines are from least squares fitting with gradient representing the sampling inverse temperature $\beta$ as in the legend.
    The Monte Carlo algorithm uses 1000 thermalization steps per sample, with 200 random initializations and $10^6$ samples per realization.
    }
    \label{fig:sampling_rrs_mv}
\end{figure}

\begin{figure}
    \centering
    \includegraphics[width=0.98\columnwidth]{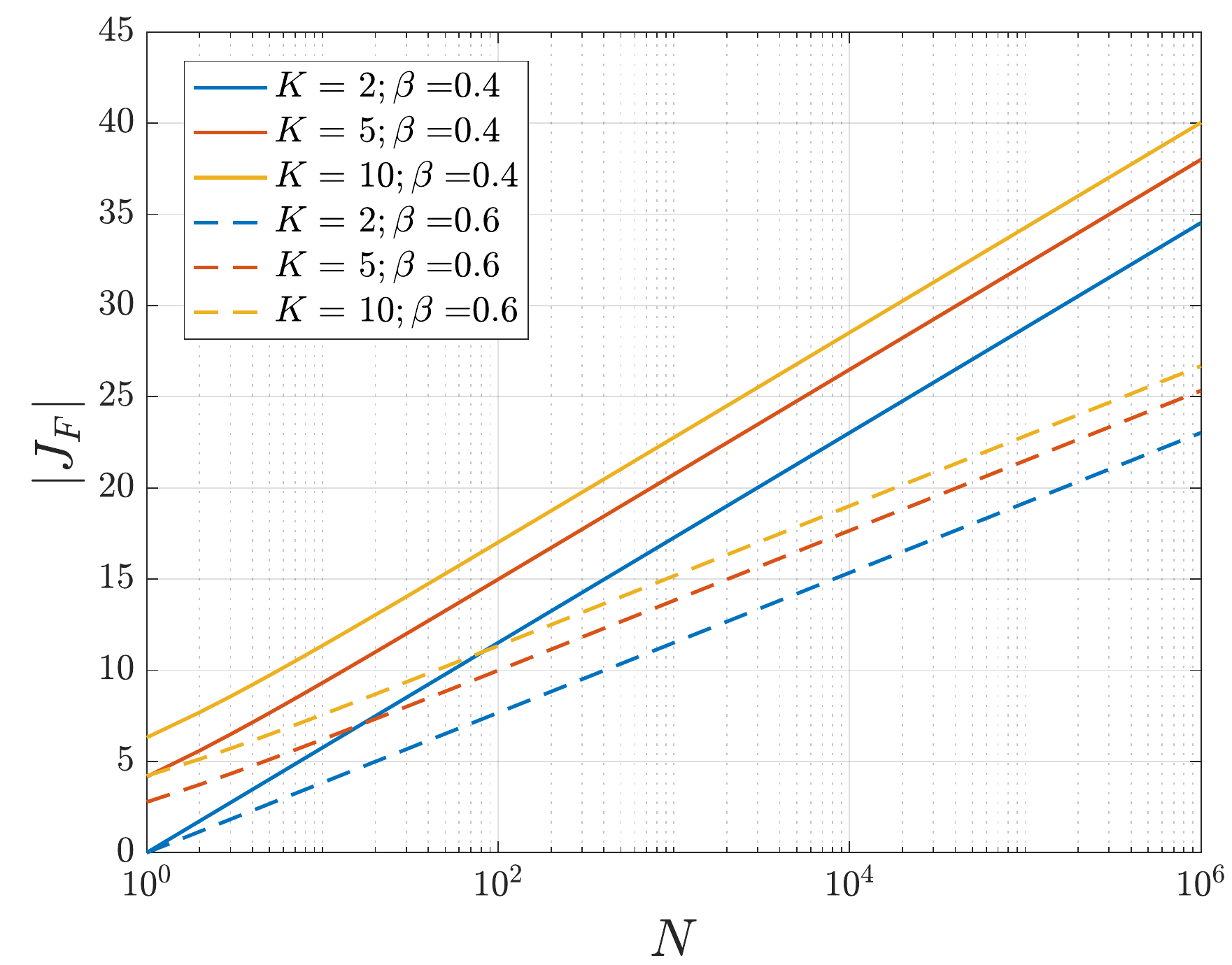}
    \caption{Graph of Eq.~(\ref{eq:Jf_scaling}) for proposed scaling of $|J_F|$  (relative to problem couplings $|J_{ij}|$) in order to minimize number of broken chains. We plot for three choices of embedding size $K$, and two temperatures.
    }
    \label{fig:Jf_scaling}
\end{figure}

One drawback of RRS is that it can be quite computationally intensive; indeed, when given a configuration where each logical spin has misaligned spins, RRS is equivalent to performing Boltzmann sampling in the entire space. If one regularly observes states where $\sim N$ logical spins are not aligned, then this will quickly become infeasible. 
By our Eq.~(\ref{eq:prob_ratio}) this is determined by the penalty weight term $\mathcal{P}_w$; since $P_n/P_{n-1}$ is decreasing in $n$ (and $P_1/P_0>1$), the most probable number $n_{\mathrm{max}}$ of broken logical spins ($P_{n_{\mathrm{max}}} \ge P_n$) is found by setting $P_n = P_{n-1}$ which gives
\begin{equation}
    n_{\mathrm{max}} = \left\lfloor \frac{N+1}{1+\mathcal{P}_w^{-1}} \right \rfloor.
\end{equation}
This means if $\mathcal{P}_w$ is `large', one may regularly find samples with $O(N)$ broken logical spins. We see therefore that in looking to sample large problem sizes would require $\mathcal{P}_w \sim O(1/N)$, which, from Eq.~(\ref{eq:penalty_weight}), can be achieved by scaling
\begin{equation}
    |J_F| \sim \frac{-1}{\beta}\log \left[\left( \frac{N+1}{N}\right)^{\frac{1}{K-1} }-1 \right].
    \label{eq:Jf_scaling}
\end{equation}
This scales very reasonably in $N$ and $K$ as shown in Fig.~\ref{fig:Jf_scaling}, suggesting the possibility of achieving this on hardware in the future.
Note that this scaling is for the absolute size of $|J_F|$, not the relative $|J_F|/|J_{ij}|$ which can be increased by reducing the $|J_{ij}|$. The distinction is that the former requires an overall increase in the energy scale available in the hardware, whereas the latter does not. See point i) in Sect.~\ref{sect:problem_embedding} for additional discussion.

However, even without this restriction there is still hope. For example, for the parameters examined in this work, if $\beta=0.6$ and $J_F=-2$ (in units of the logical Hamiltonian), for chains of length $K=3$ we get $\mathcal{P}_w = 0.19$, which means $n_{\mathrm{max}} \sim N/6.26$ for large $N$. If we wish to sample a 1000 spin (logical) problem, RRS would likely only need to handle up to 300 spins which is significantly easier.
Letting $J_f=-4$ reduces the size RRS needs to handle further to around 20 spins (with $1+\mathcal{P}_w^{-1} = 62)$.

\begin{figure}
    \centering
    \includegraphics[width=0.98\columnwidth]{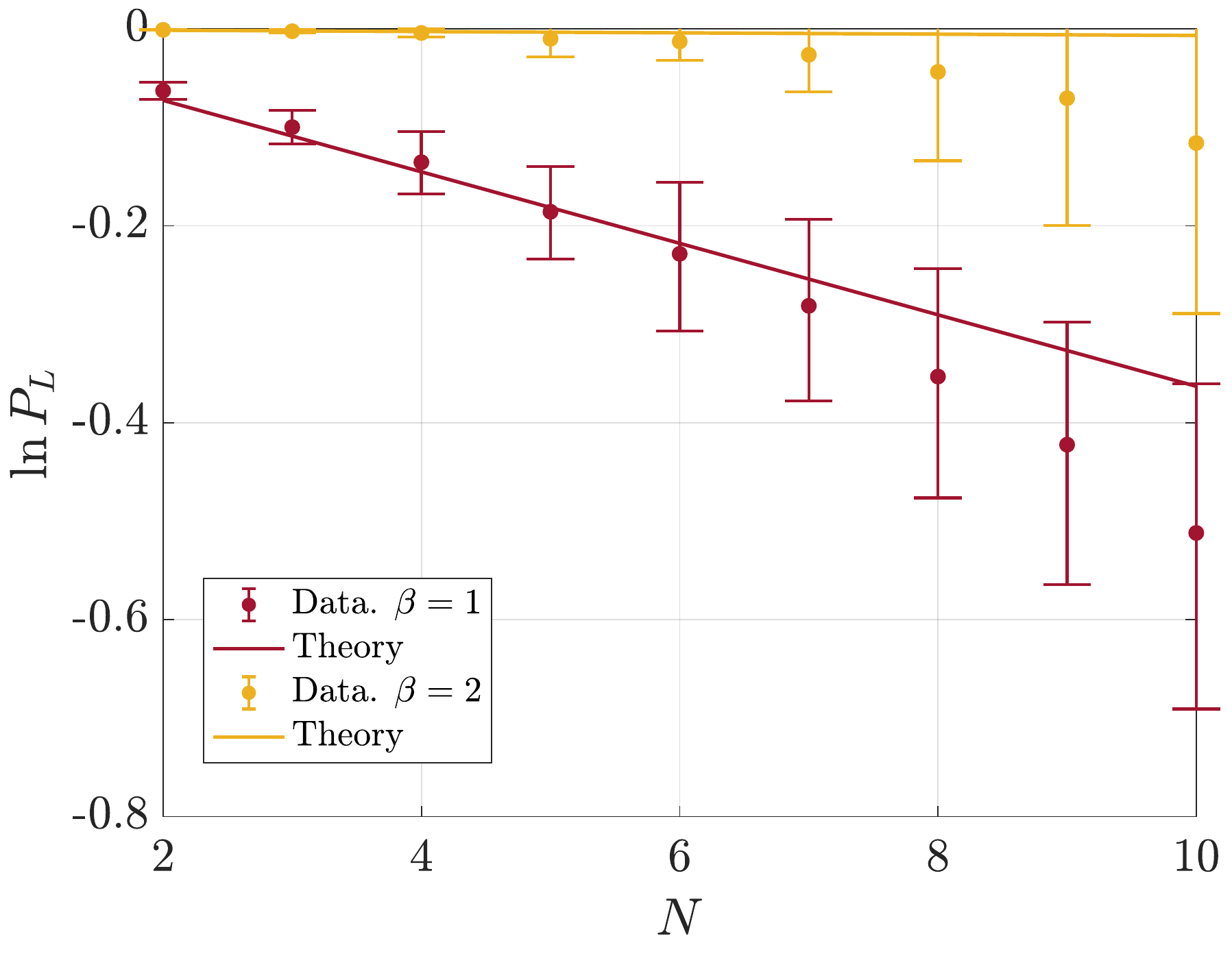}
    \caption{Same as Fig.~\ref{fig:prob_logical}, but at colder temperatures. We see our general theory seems to overestimate $P_L$.
    }
    \label{fig:prob_logical_cold}
\end{figure}

We lastly mention an interesting observation, that although our equations (\eqref{eq:prob_logical}, \eqref{eq:prob_broken2}) appear accurate for the temperatures considered here (in the average case), and are known to be accurate as $T\rightarrow 0$, for intermediate temperatures we find numerically our estimate of the logical subspace sampling $P_L$ (Eq.~\eqref{eq:prob_logical}) is in fact an overestimate (i.e. a loose, approximate upper bound). This means that relative to our derived equations, the sampling quality is in fact worse than expected, as seen in Fig.~\ref{fig:prob_logical_cold}. The reason for this may be due to the fact that our derivation assumes breaking a logical spin will always increase the energy, resulting in an underestimate of states outside of the logical subspace.
This of course means that relative to our equations, $|J_F|$ would have to scale more aggressively with $N$ than our equations predict (i.e. Eq.~\eqref{eq:Jf_scaling}). As discussed in the next section, it is worth exploring this regime in more detail, as a future research direction.

\section{Conclusion}

We have demonstrated a clear potential pitfall for any thermal sampler with a restricted topology, such as a quantum annealer for use in  understanding equilibrium physics of many body systems, phase transitions in spin glasses, and machine learning and optimization.
We showed that under the annealed approximation of spin-glasses, samples from the subspace one wishes to probe, the logical subspace, are exponentially unlikely in problem size and the complexity of the embedding (size of the logical spins $K$). 
We found analytic expressions which numerically capture this unfavourable scaling with good accuracy, for parameters studied in this work.
We proposed a new method for projecting states back to the logical subspace, and propose a scaling for the ferromagnetic coupling strength of logical spins $J_F$ which guarantees the computational plausibility of this scheme.
Fortunately, this scaling is only logarithmic in problem size $N$.

Going forward, it would be beneficial to improve, or bound in $(\beta, |J_F|)$, the accuracy of our general model (Eqs.~(\ref{eq:prob_logical}), (\ref{eq:prob_broken2})), perhaps by restricting to certain problem classes and therefore making more informed approximations.
Moreover, there are many questions about how different problem types are effected by embeddings on various topologies.
Similarly, it would be useful to obtain results for larger problem sizes and a larger range of temperatures, either analytically where possible, or through advanced sampling techniques (such as parallel tempering).
Lastly, it is clear there is a lot of room for development of new projection techniques, expanding on, or going beyond the introduced RRS scheme.
In RRS, it is assumed the temperature of the thermal sampler is known, and this may not always be the case; for example, in quantum annealers different sets of problems may be sampled at effectively different temperatures \cite{amin-freezeout,marshall-rieffel-hen-2017,powerOfPausing}. 
One would first therefore need to estimate the temperature \cite{perdomo,global_warming,wang-landau1,wang-landau2,dos_estimation}. Since in general one will not obtain the exact temperature, a further study of importance is how the performance of RRS depends on noise in the temperature parameter.

\acknowledgments
We thank Gianni Mossi, Eugeniu Plamadeala, and Max Wilson for useful discussions. 
A.~D.~G.~thanks the QuAIL group at NASA Ames and Stinger Ghaffarian Technologies, Inc.~for the kind hospitality and support while part of this work has been done.
We are grateful for support from NASA Ames Research Center. We appreciate support from the AFRL Information Directorate under grant F4HBKC4162G001 and the Office of the Director of National Intelligence (ODNI) and the Intelligence Advanced Research Projects Activity (IARPA), via IAA 145483.  The views and conclusions contained herein are those of the authors and should not be interpreted as necessarily representing the official policies or endorsements, either expressed or implied, of ODNI, IARPA, AFRL, or the U.S. Government. The U.S. Government is authorized to reproduce and distribute reprints for Governmental purpose notwithstanding any copyright annotation thereon.

\bibliography{refs.bib}

\clearpage

\appendix

\section{Average probabilities for Sect.~\ref{sect:analytic}} \label{sect:appendix_probability}
Here we explicitly obtain Eq.~\eqref{eq:prob_manychains}. 
We have $N$ chains of $K$ spins. Let us label by $\sigma_{i,\alpha}$ the $\alpha$-th spin in the $i$-th chain. The Hamiltonian is
\begin{equation}
	H = \sum_{i, j=1}^N \sum_{\alpha, \beta=1}^K J_{i\alpha}^{j\beta} \, A_{i\alpha}^{j\beta} \, \sigma_{i,\alpha} \, \sigma_{j,\beta} + J_F \sum_{i=1}^N \sum_{\alpha=1}^{K-1} \sigma_{i,\alpha} \, \sigma_{i,\alpha+1},
\end{equation}
where $J_{i\alpha}^{j \beta} = J_{j \beta}^{i\alpha}$ are the disordered interaction couplings, and $A$ is the adjacency matrix of the physical graph. We are considering the case without local fields for brevity, but the computations in that case are very similar.
Consider two configurations, $\sigma^{(\ell)}$ and $\tilde{\sigma}^{(\ell)}$,  with $\ell$ domain walls, as in the main text. We can write $\tilde{\sigma}_{i,\alpha} = \sigma_{i,\alpha} \xi_{i,\alpha}$, where $\xi_{i,\alpha}$ is 1 if the spin labeled by $i,\alpha$ has the same orientation in $\sigma^{(\ell)}$ and $\tilde{\sigma}^{(\ell)}$, -1 otherwise. To fix the ideas, we will consider, for simplicity, a bimodal distribution for the couplings (but we can immediately generalize everything to continuous distributions with zero mean and symmetric with respect to the origin):
For the disorder-averaged probability of observing $\tilde{\sigma}^{(\ell)}$ we have
\begin{widetext}
\begin{equation}
\begin{split}
	p(\tilde{\sigma}^{(\ell)}) & = \sum_{J_{i\alpha}^{j \beta} = \pm 1} \frac{1}{Z} \exp\left\{-\beta \left[ \sum_{i,j} \sum_{\alpha,\beta} J_{i\alpha}^{j \beta} \, A_{i\alpha}^{j\beta} \, \tilde{\sigma}_{i,\alpha} \, \tilde{\sigma}_{j,\beta} + J_F (N(K-1) - 2 \ell) \right] \right\}\\
	& = \sum_{J_{i\alpha}^{j \beta} = \pm 1} \frac{\exp\left\{-\beta \left[\sum_{i,j} \sum_{\alpha,\beta} J_{i\alpha}^{j \beta} \,  A_{i\alpha}^{j\beta} \, \xi_{i,\alpha} \, \sigma_{i,\alpha} \, \xi_{j,\beta} \, \sigma_{j,\beta} + J_F (N(K-1) - 2 \ell)\right] \right\}}{\sum_{\sigma_{i,\alpha}=\pm 1} \exp\left\{-\beta \left[\sum_{i,j} \sum_{\alpha,\beta} J_{i\alpha}^{j \beta} \, A_{i\alpha}^{j\beta} \, \sigma_{i,\alpha} \, \sigma_{j,\beta} + J_F \sum_{i=1}^N \sum_{\alpha=1}^{K-1} \sigma_{i,\alpha} \, \sigma_{i,\alpha+1} \right] \right\}} \\
	& = \sum_{J_{i\alpha}^{j \beta} = \pm 1} \frac{\exp\left\{-\beta \left[\sum_{i,j} \sum_{\alpha,\beta} J_{i\alpha}^{j \beta} \,  A_{i\alpha}^{j\beta} \, \sigma_{i,\alpha} \, \sigma_{j,\beta} + J_F (N(K-1) - 2 \ell) \right]\right\}}{\sum_{\sigma_{i,\alpha}=\pm 1} \exp\left\{-\beta \left[\sum_{i,j} \sum_{\alpha,\beta} J_{i\alpha}^{j \beta} \, A_{i\alpha}^{j\beta} \, \sigma_{i,\alpha} \, \sigma_{j,\beta} + J_F \sum_{i=1}^N \sum_{\alpha=1}^{K-1} \xi_{i,\alpha} \, \sigma_{i,\alpha} \, \xi_{i,\alpha+1} \, \sigma_{i,\alpha+1} \right] \right\}}\\
	& = \sum_{J_{i\alpha}^{j \beta} = \pm 1} \frac{1}{Z'} \exp\left\{-\beta \left[ \sum_{i,j} \sum_{\alpha,\beta} J_{i\alpha}^{j \beta} \, A_{i\alpha}^{j\beta} \, \sigma_{i,\alpha} \, \sigma_{j,\beta} + J_F (N(K-1) - 2 \ell) \right] \right\},
\end{split}
\end{equation}
\end{widetext}
where in the second-to-last step we have used the symmetry of the probability density function of the couplings to perform the substitution $J_{i\alpha}^{j \beta} \, \xi_{i,\alpha} \, \xi_{j,\beta} \to J_{i\alpha}^{j \beta}$ and, at the denominator, we performed the substitution $\sigma_{i,\alpha} \to \xi_{i,\alpha} \, \sigma_{i,\alpha} $. Now the numerator is the same as that of $p({\sigma}^{(\ell)})$ (Eq.~(\ref{eq:p_ell})), but the denominator is different and hence we call it $Z'$: that is we have obtained explicitly Eq.~\eqref{eq:prob_manychains}.

\section{Thermal Sampler \label{sect:appendix_sampler}}
Here we describe the Monte Carlo thermal sampler, used to generate Fig.~\ref{fig:sampling_rrs_mv}.

We implement a very basic sampler using single spin flips:
\begin{enumerate}
    \item Pick random spin configuration, compute cost $E$
    \item Pick random spin to flip, compute cost $E'$
    \item Accept change with probability $\min(1,\exp(-\beta \Delta))$  where $\Delta=E'-E$
    \item Return to step 2, and take a sample every $N_T$ steps. 
    Break after $N_S$ samples have been generated
\end{enumerate}

Here $\beta$ is the inverse sampling temperature, and $N_T$ represents a thermalization time; a sample is generated every $N_T$ steps of the algorithm. 
The total number of iterations of the above is therefore $N_T\times N_S$.

If $N_T$ is too small, samples will be heavily correlated and therefore not represent true thermal (random) samples. We typically take $N_T\approx 10\times N$ where $N$ is the number of problem variables; i.e. each spin has the chance to be flipped on average 10 times per thermalization step.
We run the steps of this algorithm over many realizations (i.e. random initial configurations) to generate statistics and to try to avoid biases such as from certain realizations becoming stuck in local minima.

\end{document}